\documentstyle[11pt]{article}
\begin{document}     
\topskip 0.0  cm     
\def\cao{\c c\~ao\ }
\def\coes{\c c\~oes\ }
\def\ii{\'\i}
\def\CAO{\c C\~AO\ }
\def\COES{\c C\~OES\ }
\def\fpe{f\'\i sica de part\'\i culas elementares\ }
\def\nao{n\~ao\ }
\def\NAO{N\~AO\ }
\def\sao{s\~ao\ } 
\def\SAO{S\~AO\ } 
\def\ao{\~ao\ }
\def\AO{\~AO\ }
\def\be{\begin{equation}\ }
\def\ee{\end{equation}}
\def\lsim{\lower.7ex\hbox{$\;\stackrel{\textstyle<}{\sim}\;$}\ }
\def\gsim{\lower.7ex\hbox{$\;\stackrel{\textstyle>}{\sim}\;$}\ }


\begin{center}
{\Large\bf  A Procura das Leis Fundamentais }
\\
(In search of fundamental laws)
\end{center}
\vskip 2 cm
\centerline{ {\bf V. Pleitez } 
}
\centerline{Instituto de F\'\i sica Te\'orica }
\centerline{Universidade Estadual Paulista }
\centerline{Rua Pamplona, 145 }
\centerline{011405-900--S\~ao Paulo, SP}
\centerline{Brazil }

\vskip 3cm
\centerline{\bf RESUMO}
Uma das atividades importantes do ensino de ci\^encias em geral, e de 
f\ii sica
em particular, \'e a discuss\ao de problemas \nao apenas atuais mas
aqueles cuja solu\cao \'e urgente. Quer dizer que deveria-se transmitir 
aos
estudantes, principalmente aos da terceira s\'erie, a imagem de uma
ci\^encia ativa, viva; deixando claro os seus sucessos e seus fracassos, 
suas dificuldades para seguir adiante.
Um ponto central dessa problem\'atica \'e a carateriza\cao do que deve 
entender-se por {\it leis fundamentais da natureza}. Em particular fazemos
\^enfase neste trabalho no fato que esse tipo de leis podem existir
em \'areas diferentes das tradicionalmente reconhecidas. Numa discuss\~ao 
dessetipo \'e imposs\ii vel (e nem mesmo desej\'avel) evitar a perspectiva 
hist\'orica do desenvolvimento cient\ii fico.

\newpage
\vskip 1cm
\centerline{\bf ABSTRACT}
One of the main activities in science teaching, and in particular in
Physics teaching, is not only the discussion of both modern problems and
problems which solution is an urgent matter. It means that the picture
of an active and alive science should be transmitted to the students,
mainly to the College students. A central point in this matter is the
issue which characterizes the Fundamental Laws of Nature. In this work
we emphasize that this sort of laws may exist in areas which are different
from those usually considered. In this type of discussion it is neither
possible nor desirable to avoid the historical perspective of the
scientific development.
\newpage

\section{ Introdu\cao} 
\label{sec:intro}
\'E frequente ouvir dizer ou ler que a ci\^encia em geral, e a f\ii sica 
em particular, est\'a 
passando por momentos dif\ii ceis. Por exemplo o n\'umero de estudantes de
f\ii sica est\'a diminuindo nos Estados Unidos~\cite{goodstein,aip99} e
provavelmente, isso ocorra no mundo todo~\cite{pf47}. No entanto 
quando analizada cuidadosamente \'e f\'acil se convencer de que a situa\cao 
n\~ao deveria ser essa.  
Paradoxalmente, isso acontece justamente no momento em que a presen\c ca
da ci\^encia \'e mais contundente na sociedade moderna~\cite{mac}. 
N\~ao d\'a para entender que n\~ao seja amplamente reconhecido que as 
contribui\coes da ci\^encia, e da f\ii sica em particular, em todos os 
aspectos da vida nas sociedades modernas t\^em sido, \sao  \'e ser\~ao 
essenciais para o desenvolvimento. 
Alguns dos problemas que a f\ii sica enfrenta, s\ao comuns
\`a ci\^encia em geral. Alguns cientistas acreditam que 
existem dificuldades nos pr\'oprios projetos da f\ii sica; outros que esses 
problemas n\ao est\ao nos temas de 
estudo da f\ii sica mas nas suas rela\coes com a sociedade~\cite{mac}. 
De qualquer forma,
a vis\~ao pesimista sobre as \'areas de pesquisa na f\ii sica enfraquece as
rela\c c\~oes dela com a sociedade. Assim, discutir em n\ii vel estrictamente
cient\ii fico quais os rumos e dificulades da f\ii sica, contribui a 
melhorar o di\'alogo com a sociedade. Assim podemos perguntar-nos  
o que seria necess\'ario fazer para manter em bom estado 
a pesquisa, o ensino e a influ\^encia cultural da ci\^encia em geral, e da 
f\ii sica em particular?~\cite{mac,schat}. 

\'E necess\'ario que, entre outras coisas, se
fa\c ca \^enfase na import\^ancia da ci\^encia f\ii sica 
b\'asica~\footnote{Por ``ci\^encias f\ii sicas'' entendemos a totalidade 
das ci\^encias
f\ii sicas: astronomia, astrof\ii sica, cosmologia, materia condensada,
f\ii sica do meio ambiente, f\ii sica de part\ii culas elementares, etc.}  
de maneira que se promova a f\ii sica orientada, motivada pela curiosidade; 
temos que reconhecer a import\^ancia de educar e informar ao p\'ublico, 
isto \'e, a divulga\c c\~ao cient\ii fica; tamb\'em em melhorar o ensino da 
f\ii sica e o jornalismo cient\'\i fico. 
Apenas as motiva\coes de curto prazo e econ\^omicas n\ao s\ao suficientes. 
Devemos sempre ressaltar que os conceitos f\ii sicos s\ao a base dos 
microprocessadores, do laser e da fibra \'otica; somente para mencionar
algumas das contribui\coes importantes baseadas em princ\ii pios b\'asicos. 
Por\'em poderiamos retroceder at\'e o s\'eculo passado e mencionar
muitas outras contribui\c c\~oes da ci\^encia ou ent\~ao tentar prever quais 
ser\ao os futuros impactos quando
as revolu\coes do minilaser~\cite{gourley}, da computa\cao qu\^antica  
sejam realidade~\cite{compu} ou, mesmo os avan\c cos imprevis\'\i veis em 
outras \'areas como as Ci\^encias da Terra: ou ser\'a que descobrir qual \'e 
o mecanismo respons\'avel pelo movimento das placas tect\^onicas n\ao \'e 
fundamental? acaso n\~ao ter\'a impacto no desenvolvimento 
futuro conhecer melhor a evolu\c c\~ao interna da Terra?~\cite{maddox}. 

Acreditamos que uma discus\~ao sobre o que \'e a procura de {\it leis 
fundamentais da natureza} possa contribuir um pouco para o esclarescimento 
dessa problem\'atica complicada.
Afinal, a curiosidade continuar\'a a ser uma motiva\c c\~ao para alguns
estudantes seguirem uma carreira cient\ii fica, a f\'\i sica por exemplo.

\subsection{Un pouco de hist\'oria}
\label{subsec:hsitoria}

Pode-se dizer que, em certo sentido, a f\ii sica contempor\^anea 
come\c cou com Cop\'ernico, Galileo e outros~\cite{cohen}. Por outro lado, 
a primeira s\ii ntese conceitual na descri\cao dos fen\^omenos observados
na \'epoca, foi a de Newton no s\'eculo XVII. As leis de Newton e outros 
princ\ii pios gerais, como as leis de conserva\c c\~ao, permitiram a 
descri\cao de todos os fen\^omenos conhecidos at\'e a \'epoca de Newton e nos 
anos posteriores.~\footnote{Havia algumas discrep\^ancias mas, para a
exposi\cao sucinta que estamos fazendo isto \nao tem import\^ancia.
Isto nos levaria a considerar a quest\ao delicada de quando um
experimento \'e crucial~\cite{pl99}.}
Quando afirmamos que a teoria de Newton foi uma {\em s\ii ntese} queremos
dizer que ela permitiu que processos aparentemente \nao
relacionados, como o movimento dos planetas e os observados aqui na
Terra, fossem descritos por um \'unico conjunto de princ\ii pios.

No s\'eculo seguinte (Sec. XVIII) foi realizado o desenvolvimento 
ma\-te\-m\'a\-tico 
da mec\^anica cl\'assica newtoniana. Hamilton, Lagrange e outros nomes bem 
conhecidos. Quase que concomitantemente, no come\c co do s\'eculo XIX, foram
feitos uma s\'erie de experimentos sobre fen\^omenos el\'etricos e 
magn\'eticos que culminaram com a descoberta, por Faraday, Amp\`ere e 
ou\-tros pesquisadores, das leis da eletricidade e do magnetismo as quais logo 
seriam {\it unificadas}, junto com a \'otica, na teoria do campo 
eletromagn\'etico de Maxwell. (Esta foi a segunda grande {\it s\'\i ntese} 
nas leis dos fen\^omenos naturais.)

Temos ent\ao que no come\c co do s\'eculo XX, uma din\^amica de 
part\'\i culas relativ\'\i stica (Einstein) e a eletrodin\^amica de Maxwell 
(tamb\'em relativ\'\i stica)  
formabam os pilares do nosso co\-nhe\-ci\-men\-to cient\ii fico das leis 
b\'asicas da natureza. Essas teorias constituem o que se conhe\c ce hoje pelo 
nome de {\it F\ii sica Cl\'assica}. 

No final do s\'eculo passado ainda a exist\^encia dos \^atomos \nao era 
amplamente aceita, ou seja que n\~ao se acreditava que a {\sl microf\ii sica}
fosse cons\-ti\-tu\ii\-da de fen\^omenos diversos dos observados em escalas 
macrosc\'opicas. Apenas em 1913 as experi\^encias de J. Perrin mostraram que
os \'atomos, os quais os qu\ii micos usavam apenas como uma maneira de
des\-cri\-\cao das rea\c c\~oes qu\ii micas, tinham exist\^encia 
real~\cite{nye}. Tampouco havia 
nessa \'epoca uma vi\sao do universo como um todo, isto \'e, 
o conceito de que o universo evolue a partir de um estado 
inicial.~\footnote{
\'E interessante observar que, ainda que muitas das id\'eias
na f\ii sica moderna tenham, de alguma forma, um conceito an\'alogo na Grecia
antiga, este n\~ao \'e o caso de um universo em expans\~ao. Este \'ultimo
\'e um conceito que nasce no nosso s\'eculo.}

Por outro lado, o chamado {\it problema do corpo negro}, isto \'e, a lei que 
descreve a intera\cao da radia\cao em equilibrio t\'ermico com a mat\'eria,
estava em aberto, e os experimentos \nao confirmavam os
modelos te\'oricos para explicar esse fen\^omeno. A resolu\cao  do problema 
levaria, no percurso das d\'ecadas seguintes, \`a descoberta da {\it 
mec\^anica qu\^antica}, a teoria que substitui a mec\^anica de Newton 
no caso de fen\^omenos na escala at\^omica (da ordem de $10^{-8}$ cm).

Entre 1895 e 1897, foram feitas 3 descobertas experimentais que teriam grandes 
implica\coes ao longo de todo o s\'eculo XX:
\begin{itemize}
\item a descoberta dos Raios-X por R\"ontgen,
\item a descoberta da radioactividade natural por Becquerel,
\item a descoberta do el\'etron por J. J. Thompson.
\end{itemize}

As duas primeiras foram feitas por acaso. Nos anos seguintes ficaria
claro que os raios-X \sao ondas eletromagn\'eticas de grande energia e 
que a radio\-a\-ti\-vi\-da\-de era um fen\^omeno at\^omico ou, melhor, 
nuclear. Isto \nao era evidente mas foi demostrado por Rutherford nas
primeiras d\'ecadas deste s\'eculo. A descoberta de Thompson e outras
experi\^encias posteriores, mostraram
que os portadores da eletricidade negativa \sao {\em constituentes 
universais} da mat\'eria. Estava assim, descoberta a primeira part\ii cula 
elementar~\cite{pr}.

Podemos dizer, de maneira resumida, que os f\'\i sicos no come\c co do 
nosso s\'eculo estudavam experimentalmente a radioactividade, os te\'oricos 
pro\-pu\-nham modelos do \'atomo, outros pesquisadores experimentais 
estudavam os raios
c\'osmicos ou tentavam obter baixas temperaturas. Te\'oricos
como Einstein estudavam a generaliza\cao da relatividade restrita
e que o levaria \`a proposta da relatividade geral.

At\'e o come\c co dos anos 30 pensava-se que todos os fen\^omenos
naturais tinham origem em apenas duas for\c cas fundamentais: a
gravita\c c\~ao e a eletromagn\'etica. Estas teorias eram descritas como
campos cl\'assicos preenchendo o espa\c co todo. As suas fontes eram a
massa e a carga el\'etrica, respectivamente. No caso gravitacional as
equa\coes de Einstein descrevem a gravita\cao em condi\coes especiais,
mas a teoria de Newton \'e usada na maioria das aplica\coes do
dia-a-dia. 

Pouco tempo depois, ainda nos anos 30, foi reconhecido que para
explicar fen\^omenos at\^omicos e  sub-at\^omicos (nucleares) era 
necess\'ario admitir a e\-xis\-t\^en\-cia de duas outras for\c cas: a 
fraca e a forte. A primeira, a for\c ca fraca, para explicar o decaimento 
ra\-dio\-a\-ti\-vo $\beta$ e a segunda, para garantir a estrutura nuclear. 
Nenhuma das duas for\c cas \'e observada macrosc\'opicamente e, 
contrariamente \`as for\c cas gravitacionais e eletromagn\'eticas, devem 
ter alcance muito curto. 

At\'e hoje, as 4 for\c cas podem ser tratadas
separadamente. Em termos observacionais, isso significa quatro escalas
diferentes para as se\coes de choque e vidas m\'edias dos diferentes
processos entre as part\ii culas elementares at\'e agora observadas. 
A descri\cao atual das for\c cas fracas e fortes est\'a baseada em
teorias de calibre (ou de {\it gauge}) locais que t\^em como exemplo a 
e\-le\-tro\-di\-n\^a\-mi\-ca qu\^antica (QED).
Todo este esquema n\ao foi obtido sem reservas. Afinal a ci\^encia tem
de ser c\'etica e o preconceito, sejam positivos (adiantam o reconhecimento
de um fato ou teoria) ou negativos (dificultam o mesmo) mora ao lado.

\subsection{Motiva\c c\~oes}
\label{subsec:enfase}
Acreditamos que uma discuss\ao sobre o que poderia ser a procura de 
{\em leis fundamentais} ajudar\'a a estudantes de pos-gradua\cao na 
escolha ou na va\-lo\-ri\-za\-\cao das suas respectivas \'areas de pesquisa 
e aos da gradua\cao a escolher sua futura \'area de trabalho. 
Se o n\'umero de estudantes de F\'\i sica esta diminuindo, 
como poderia se reverter essa tend\^encia? Qualquer que seja a resposta
a este desafio uma das suas componentes ser\'a a motiva\cao dos estudantes 
sobre o que \'e importante pesquisar. 
Ent\~ao, se faz necess\'ario uma discuss\~ao
sobre onde e como podemos procurar esse tipo de leis fundamentais. 
Este \'e um ponto importante e esperamos que este
artigo possa contribuir, ainda que modestamente, a repensar o assunto.
Sim, repens\'a-lo porque j\'a existe uma resposta tradicional \`a pergunta
de onde podemos identificar as leis fundamentais. No momento que novos fatos
ou propriedades da mat\'eria s\ao descobertos, essa resposta n\~ao \'e
mais apropriada. Precissamos ent\ao redescobrir qual o sentido das leis 
fundamentais. 

Quer dizer que no ensino de ci\^encias os aspectos pedag\'ogicos n\ao s\ao 
mais suficientes. Se ensinar o que sabemos \'e dif\ii cil, n\ao o \'e menos 
ensinar o que n\ao sabemos. 
N\ao saber no sentido amplo do termo: coisas que a ci\^encia est\'a ainda 
pesquisando ou mesmo n\ao tem condi\c c\~oes, no momento, de responder. 

Existem outros aspectos do problema como a educa\cao do p\'ublico em geral.
Convencer \`as pessoas que a f\ii sica continuar\'a a ser a base da ci\^encia 
e a tecnologia no futuro y que tambi\'en jogar\'a un papel importante
na an\'alise e resolu\cao de problemas energ\'eticos e do meio ambiente.
Mas antes de chegar ao p\'ublico, precissamos convencer os estudantes
sobre quais s\ao os problemas fundamentais que devem ser atacados por
eles. Que existem problemas fundamentais em \'areas n\ao reconhecidas
por uma mentalidade infantil que infelizmente ainda permeia os nossos meios 
acad\^emicos. Um aspecto que n\ao ser\'a tratados aqui \'e o fato que as
diretrizes metodol\'ogicas n\ao s\ao suficientes para caraterizar a atividade 
cient\ii fica~\cite{pl96,pl99}. 

\section{Rompendo barreiras}
\label{sec:2}
O m\'etodo cient\ii fico, qualquer coisa que entendamos por isso, \nao
tem um ant\ii doto contra os preconceitos. Por exemplo, mesmo no
come\c co do presente s\'eculo f\ii sicos como Lord Kelvin (e Mach como 
veremos mais adiante) nao acreditavam na exist\^encia dos \'atomos. 
Segundo eles os \'atomos seriam apenas
abstra\coes \'uteis para os qu\ii micos. No entanto, o mesmo Lord
Kelvin escreveu no pref\'acio do livro de Hertz~\cite{hertz}
\begin{quotation}
{\em The explanation of the motion of the planets by a
mechanism of crystal cycles and epicyles seemed natural and
intelligible, and the improvement of this mechanism invented by
Descartes in his {\em vortices} was no doubt quite satisfactory to
some of the greatest of Newton's scientific contemporaries.
Descartes's doctrine died hard among the mathematicians and
philosophers of continental Europe; and for the first quarter of
last century belief in universal gravitation was insularity of our
countrymen.  }
\end{quotation}

Segundo Weinberg ~\cite{sw2}
\begin{quotation}
{\em The heroic past of mechanism gave it such prestige that the 
followers of Descartes had trouble accepting Newton's theory of the 
solar system. How could a good Cartesian, believing that all natural 
phenomena could be reduced to the impact of material bodies or fluids 
on one another, accept Newton's view that  the sun exerts a force on 
the earth across 93 million miles of empty space? It was not until 
well the eighteen century that Continental philosophers began to feel 
comfortable with the idea of action at a distance. In the end Newton's 
ideas did prevail on the Continent as well as in Britain, in Holland,
Italy, France, and Germany (in that order) from 1720 on. }
\end{quotation}
Apenas em 1728 ap\'os uma viagem de Voltaire a Londres a escola
Newtoniana come\c cou a ter disc\ii pulos em Paris~\cite{west}. 
N\~ao \'e surpreendente que o conceito de {\it a\c c\~ao a dist\^ancia} 
\nao era aceito pela comunidade. 
\'E interessante que o pr\'oprio Newton disse~\cite{whita}
\begin{quotation}
{\em ...that one body  may act upon another at a distance through vacuum,
without the mediation of anything else ... is to me so great absurdity,
that I believed no man, who has in philosophical matters a competent 
faculty for thinking, can ever fall into.}
\end{quotation} 

Por alguns anos, depois de 1687 (ano da publica\cao dos {\it Principia}), 
mesmo em Cambridge, continuo-se a ensinar o Cartesianismo. Apenas ocorreu 
que no Continente as ideais de Newton demoraram um pouco mais para serem 
aceitas~\cite{whita}. Voltaire escrevia em 1730~\cite{whita}
\begin{quotation}
{\em  A Frenchman who arrives in London will find philosophy, like 
everything else, very much changed there. He has left the wordls a 
{\it plenum}, and now he find a {\it vacuum}.  It is the language 
used, and not the thing in itself, that irritates the humand mind. If 
Newton had not used the world {\it attraction} in his admirable 
philosophy, every one in our Academy would have open his eyes to the 
light; but unfortunately he used in London a word to which an idea of 
ridicule was attached in Paris...}
\end{quotation}

Segundo Whittaker~\cite{whita}
\begin{quotation}
{\em In Germany, Leibnitz described the Newton formula as a return to 
the disacredited scholastic concept of {\it occult qualities} and a 
late as the middle eighteenth century Euler and two of the Bernoullis 
based the explanation of magnetism on the hypothesis of vortices.}
\end{quotation}
Deve-se lembrar tamb\'em que essa oposi\cao entre disc\ii pulos de Newton 
e Descartes fez que os primeiros rejeitassem, posteriormente, a id\'eia 
de {\em \'eter} nos fen\^omenos el\'etricos e magn\'eticos. Vemos que como
dissemos antes, o preconceito mora ao lado, a verdade aparece sempre com 
dificuldades! Neste caso o curioso \'e que posteriormente a vis\ao 
newtoniana passou a ser o preconceito contra novas formas de descrever o 
mundo f\ii sico.

\section{Desafios}
\label{sec:sec2}
Usualmente, descreve-se o desenvolvimento da f\ii sica 
como a evolu\cao da explica\cao de fen\^omenos relativos a uma
determinada escala das dimens\~oes espaciais e do tempo, em termos de
processos mais elementares caracter\ii sticos de uma
escala espa\c co-temporal menor. Foi o que aconteceu com a
descoberta da estrutura at\^omica da mat\'eria a qual sabemos agora que
\'e composta de \'atomos e mol\'eculas. Logo se constatou que os \'atomos
por sua vez \sao cons\-ti\-tu\-i\-dos por el\'etrons e pelo n\'ucleo 
at\^omico. Este
\'ultimo \'e formado pelos n\'ucleons que por sua vez \sao formados
pelos quarks. Foi esta hierarquia de fen\^omenos que levou os cientistas a 
acreditar que as leis fundamentais eram apenas aquelas que permitiam descer
na escala das dimens\~oes espaciais e do tempo. Isto \'e, o desenvolvimento
da ci\^encia, e em particular o da f\'\i sica, foi at\'e pouco tempo
totalmente {\sl reducionista}. 

At\'e onde vai esta cadeia? Sem d\'uvida a resposta a
esta pergunta faz parte da chamada {\it pesquisa b\'asica}. Por\'em, este
tipo pesquisa est\'a restrita \`a procura de novas leis carater\ii sticas de
escalas menores? N\ao h\'a novas {\em leis fundamentais},
por exemplo, na escala humana ou a n\ii vel at\^omico? Se a resposta for 
positiva, como podemos reconhecer leis fundamentais? Se for negativa, por 
qu\^e ?  
N\ao \'e f\'acil uma defini\cao de {\it lei fundamental}. De fato, nenhum
defini\cao \'e f\'acil. Mas podemos reconhec\'e-la. Quando um conceito ou lei
n\ao depende de outro de maneira direta que o explica podemos dizer que
o primeiro \'e um conceito ou lei fundamental. Assim, a qu\ii mica tem
conceitos e leis que n\ao podem ser reduzidos \`a f\ii sica. Isto \'e,
a qu\ii mica tem seu estatus particular como ci\^encia da natureza mesmo que
seus fundamentos estejam baseados nas leis da f\ii sica. Mas  existem ainda
mesmo \'areas da f\ii sica onde as leis cl\'assicas ou qu\^anticas ajudam 
pouco para  se estabelecer suas leis e conceitos. Um exemplo, a ser discutido
 mais adiante, \'e o {\it caos 
determin\ii stico}.  

Al\'em das dificuldades intr\ii nsecas, a resposta \`a pergunta acima 
\'e par\-ti\-cu\-lar\-men\-te delicada, porque a situa\cao atual da 
f\ii sica te\'orica \'e, em certo sentido, de crise. 
As palavras recentes de Schweber resumem a problem\'atica 
atual~\cite{sss} 
\begin{quotation}
{\em A deep sense of unease permeates the physical
sciences...Tra\-di\-tio\-nally, physics have been highly reductionist,
analyzing nature in terms of smaller and smaller building blocks and
revealing underlying, unifying fundamental laws...Now, however, the
reductionist approach that has been the hallmark of theoretical
physics in the 20th century is being superseded by the investigation
of emergent phenomena, the study of the properties of complexes whose
`elementary' constituents and their interactions are known.
Physics, it coul be said, is becoming like chemistry.}
\end{quotation}

As pesquisas cient\ii ficas \sao divididas segundo Weisskopf 
em {\it intensivas} e {\it extensivas}. As do primeiro tipo teriam a ver 
com a procura de leis fundamentais, as do segundo tentam descrever
os fen\^omenos em termos das leis fundamentais conhecidas~\cite{vw2}.
Neste sentido a f\ii sica da mat\'eria condensada, f\ii sica de
plasma e outras \'areas seriam do tipo extensivo, entanto que a 
f\ii sica 
de altas energias e parte da f\ii sica nuclear seriam intensivas.

Tomada literalmente \'e uma maneira de desenvolvimento ``barroca'', 
isto \'e, 
uma disciplina \'e separada numa multid\ao de \'areas, 
uma quantidade de detalhes e complexidades desorganizados. 
Isto pode ocorrer em ci\^encias matematizadas ou mesmo nas ci\^encias 
emp\'\i ricas.

Na verdade estamos numa \'epoca de grandes mudan\c cas em que as 
a\-ssun\-\coes b\'asicas da pesquisa nas diversas \'areas da f\'\i sica
parecem deslocadas com rela\cao as anteriores: a {\em complexidade} e 
a {\em emerg\^encia} (o da turbul\^encia por exemplo) parecem ser os 
objetivos principais a serem tratados~\cite{sss}.
Outra \'area de grande futuro s\~ao as t\'ecnicas de \'optica qu\^antica.
\'E poss\ii vel prever at\'e onde nos levara os novos testes dos 
princ\'\i pios da mec\^anica qu\^antica?
Desde os experimentos de Aspect e colaboradores~\cite{aspect} que testaram
as desigualdades de Bell e mostraram que a interpreta\cao ortodoxa era
confirmada, pasando pelos efeitos ``superluminares'' de Chiao {\it et 
al.}~\cite{steinberg,chiao}, at\'e testes mais recentes~\cite{ghose}, 
indicam que podemos estar assistindo \`a des\-co\-ber\-ta de novos 
fen\^omenos qu\^anticos e
isso ter\'a importantes conseq\"u\^encias em computa\c c\~ao (que cada vez
est\'a atingindo dist\^ancias menores) e, por isso,
em  todas as  outras \'areas da ci\^encia e da tecnologia. Tudo isso n\~ao
parece t\~ao fundamental e b\'asico como outras leis da natureza?

Paradoxalmente, a situa\c c\~ao, no caso da f\ii sica de part\ii culas 
elementares, \'e uma consequ\^encia do sucesso 
da teoria qu\^antica de campos e do uso das simetrias, locais e globais.
fica dif\'\i cil de prever qual ser\'a o formalismo que substituir\'a
ao atual. 
No entanto, quando apropriadamente 
considerada, a situa\cao atual \'e empolgante. 
Acreditamos apenas que a f\'\i sica entrou numa nova fase de maturidade 
nas diversas \'areas. Por exemplo, a f\'\i sica de neutrinos est\'a numa 
fase de muita coleta de dados experimentais dos quais poder\'a sair dados
definitivos das propiedades dos neutrinos~\cite{nus,gefan}. 

O sentimento de dificuldade acima mencionado, \nao ocorre apenas na 
f\ii sica de part\ii culas elementares. 
O mesmo ocorre em \'areas como 
a mat\'eria condensada e a cosmologia, mesmo (ou justamente por causa 
deles) com os dados recentes do COBE~\cite{cobe}), parecem estar numa 
situa\cao de aparente falta de perspectivas. No caso da mat\'eria 
condensada \nao tem havido avan\c cos na compreen\sao dos fen\^omenos 
cr\ii ticos e a superconditividade a altas temperaturas ainda \nao tem 
uma teoria bem estabelecida~\cite{sss}. Mas podemos assinalar para a 
descobertas experimentais da condensa\c c\~ao de Bose-Einstein com 
diversos tipos de \^atomos, inclusive o hidrog\^enio~\cite{cornell,bec}. 
Mesmo em \'areas de grande desenvolvimento
recente como o {\em caos determin\ii stico} e fen\^omenos relacionados, 
parece ter-se alcan\c cado uma estabilidade nas descobertas te\'oricas e 
ex\-pe\-ri\-men\-tais~\cite{ruelle}.  
Os {\em fractais} tampouco produziram uma renova\cao da nossa vi\sao da 
natureza (pelo menos por enquanto) e servem (quase) apenas para produzir 
figuras ex\'oticas com ajuda de computadores~\cite{leo}. A \'area da 
programa\c c\~ao, a despeito dos grandes avan\c cos, continua na sua crise 
peremne~\cite{wwg}.    

Tudo isso est\'a relacionado com o que esperamos
da f\ii sica como um todo e, em particular, da f\ii sica te\'orica.
Considero que transmitir esse tipo de {\it ansiedade} \'e fundamental
no {\it ensino de f\'\i sica}. Precissamos ensinar n\~ao apenas o conhecido
mas tamb\'em o desconhecido, o que est\'a sendo pesquisado no momento pelos
especialistas das diversas \'areas. Deve-se fazer \^enfase na ignor\^ancia
da ci\^encia em certos assuntos. Isso coloca a prioridade da atualiza\cao
dos professores com rela\cao \`as necessidades puramente pedag\'ogicas. 
Mais \nao apenas isso. Na atualidade a vida das pessoas
\'e cada vez mais afetada pela ci\^encia e a t\'ecnica. Elas precissam 
entender melhor em que consiste o m\'etodo cient\'\i fico ou melhor, em que 
consiste a {\it maneira cient\'\i fica} de pensar e agir (e tamb\'em quais 
\sao as suas limita\c c\~oes). Essa necessidade \'e fazer `compreender' a 
ci\^encia pelos estudantes (e o p\'ublico geral) \'e mais importante que 
apenas a mera atualiza\c c\~ao dos resultados obtidos pela ci\^encia e a 
tecnologia.

Do ponto de vista dos pr\'opios pesquisadores e dos estudantes de 
p\'os-gradua\c c\~ao as medita\c c\~oes s\~ao mais delicadas, mas nem por 
isso menos urgentes ou necess\'arias. \'E urgente e/ou necess\'ario obter 
uma fun\cao de onda para o universo (mesmo que o universo primordial)? 
pode-se obter uma teoria de tudo (``theory of everything'') com os 
conhecimentos emp\'\i ricos atuais? A prioriza\cao dos 
objetivos da pesquisa \'e essencialmente uma escolha pessoal, ainda que 
outros fatores influenciem nela (como o financiamento, mercado de trabalho, 
a influ\^encia do orientador na pos-graduac\~ao). 
Uma interrogante importante sempre ser\'a sobre o qu\^e estamos em 
capacidade de verificar experimentalmente. A especula\cao \'e valida mas
temos de ter cuidado em \nao cair numa situa\cao {\it grega}, isto \'e, uma 
situa\cao onde apenas o conceito de teoria matematicamente ``bela'' \'e o 
que importa. Esse conceito \'e certamente relativo. 

Nesse quadro geral, o problema \'e colocado aos pesquisadores e, em 
particular aos estudantes que come\c cam sua p\'os-gradua\c c\~ao, de 
escolher rumos nas suas pesquisas. A escolha \'e certamente um assunto 
pessoal. Todos est\ao sozinhos ao faz\^e-la. Vale a pena, no entanto, 
fazer an\'alises que possam, pelo menos, colocar o assunto em discuss\~ao 
de maneira que v\'arios crit\'erios possam ser levados em conta na hora de 
escolher.  

Um aspecto que atrai os pesquisadores para determinados campos da 
pesquisa \'e o fato de ela ser considerada ampla e ``fundamental''. 
Isto \'e, base de tudo o resto, que seria constitu\ii do apenas de 
detalhes. Os conceitos de ``import\^ancia'', ``beleza'' e 
``consist\^encia'' \sao 
tamb\'em, frequentemente trazidos \`a tona. 

Se uma \'area \'e ``fundamental''  ou, aceitando que essa palavra seja 
sin\^onima de ``importante'', ent\ao ela deve ser relevante para 
\'areas vizinhas. Por exemplo, parece indiscut\ii vel que h\'a varios 
anos a biologia molecular \'e a \'area mais fundamental das ci\^encias 
biol\'ogicas.
Assim, um estudante pode ser motivado a escolher essa \'area de pesquisa. 
Os objetivos dessa ci\^encia (compreender melhor a transmiss\ao da 
informa\cao gen\'etica) s\~ao, aparentemente, mais f\'aceis de identificar. 
Sua import\^ancia com rela\cao a doen\c cas como c\'ancer, aids e outras, 
assim
como a sua utiliza\cao em t\'ecnica recentes de produtos transg\^enicos e
clonagens \'e indiscut\ii vel. 
Se a f\ii sica de altas energias \'e vista como uma maneira 
de entender melhor as for\c cas nucleares ent\ao poderia ser comparada com 
a biologia molecular. 
No entanto, esse objetivo foi deixado de lado, no que se refere aos fatos 
principais a serem explicados, e se procura uma 
unifica\cao das for\c cas observadas (at\'e o momento) na natureza. 
Ainda que isto possa ser uma motiva\cao para atrair jovens talentosos, 
poderia ser uma maneira, a curto prazo, de frustr\'a-los e perder quadros 
valiosos. 

Em 1964 Alan Weinberg~\cite{aw} observara que o afastamento da 
f\ii sica de altas energias do resto das outras \'areas da f\ii sica 
diminue a sua importancia como ci\^encia fundamental. 
Claro que como ci\^encia tem objetivos bem definidos e ambiciosos. 
O problema, \'e que \'e cara. Por isso suas verbas \sao cada vez mais 
dif\ii ceis de serem obtidas nos paises do primeiro mundo. Em parte porque 
tem de competir com \'areas e/ou temas de pesquisa novos, isto \'e, que 
n\~ao existiam 10 ou 15 anos atr\'as (pelo menos n\~ao de maneira 
estruturada).
Por outro lado, devemos lembrar que a ci\^encia \'e uma s\'o. 
Assim, se um projeto \'e cancelado no primeiro mundo vai nos afetar 
tamb\'em. 
Nos \nao podemos ficar, pelo menos na \'area  de f\ii sica te\'orica, 
resolvendo problemas diferentes dos da comunidade internacional. 
N\ao devemos aceitar a divi\sao do trabalho internacional. 
O desafio, levando em considera\cao as diferen\c cas de recursos, 
\'e o mesmo.

A ``beleza'' e a ``consist\^encia'' \sao fatores muitas vezes mais
determinantes que a observa\cao experimental na aceita\cao de uma 
determinada teoria. Isso serve para decidir entre duas teorias com 
diferentes graus de 
beleza. Mas, este adjetivo tem unicidade? 
{\it i.e.}, podemos formular apenas ``uma'' \'unica teoria bela? Pelo 
menos por enquanto este parece ser o caso da Relativiadade Geral e da 
Mec\^anica Qu\^antica. O caso desta \'ultima \'e mais impressionante. 
Podemos fazer corre\coes \`a Relatividade Geral acrescentando termos 
\` a lagrangeana mas \nao sabemos como modificar apenas ``um pouco'' a 
mec\^anica qu\^antica!~\footnote{Isto \'e, podemos sim modificar as 
rela\c c\~oes de comuta\c c\~ao.} Por outro lado, esta \nao determina o 
tipo de part\ii culas e suas intera\c c\~oes.

A criterio de beleza sempre foi utilizado pelos cientistas. Segundo 
Chandrasehkar~\cite{chan}
\begin{quotation}
{\em Science, like arts, admits aesthetic criteria; we seek theories that 
display  a proper conformity of the parts to one another and to the whole
while still showing {\it some strange in their proportion}.}
\end{quotation}

O problema \'e que mesmo na arte o criterio de beleza \'e cultural e 
depende tamb\'em do tempo. 
Pior, na ci\^encia como na arte os preconceitos t\^em um papel, para bem 
o para mal, importante. A beleza manifesta, para n\'os, da teoria 
at\^omica \nao era evidente para grandes f\ii sicos de s\'eculo pasado, 
como Lord Kelvin e outros. Mach por exemplo dizia que~\cite{sw2}:
\begin{quotation}
{\em If believed in the reality of atoms is so crucial, then I renounce 
the physicsl way of thinking. I will not be a professional physicist, and I 
hand back my scientific reputation.}
\end{quotation}

Qualquer que seja a defini\cao de beleza para teorias cient\ii ficas, a 
``simplicidade'' deve fazer parte dela. Mas, como se mede a simplicidade? 
Segundo Weinberg~\cite{sw2}, \'e a simplicidade de id\'eias o que importa. 
Rubbia \'e mais enf\'atico:  o ``script'' \'e mais importante que os 
``atores''. A teoria de Newton \'e constituida por 3 equa\coes entanto que 
a do Einstein tem 10!
Mas sem d\'uvida nenhuma a \'ultima \'e considerada, pela maioria da 
comunidade de f\ii sicos, como sendo mais bela (e fundamental) que a 
primeira! Assim, \nao devemos identificar a simplicidade con o n\'umero 
m\ii nimo de qualquer coisa. \'E interessante que o chamado ``modelo 
padr\~ao'' das \fpe a despeito 
de contar com um n\'umero grande de par\^ametros \'e de uma grande 
simplicidade na descric\ao das intera\coes entre part\ii culas 
e\-le\-men\-ta\-res. E, o que \'e mais importante, o modelo \nao depende 
fortemente dos valores que esses par\^ametros venham a ter na realidade. 
Na eletrodin\^amica cl\'assica, alguns dos par\^ametros como o \ii ndice 
de refra\cao tem que ser obtidos experimentalmente. Isso \nao tira beleza 
\`a teoria de Maxwell.

Por outro lado, \'e bom frisar que a explica\cao de porqu\^e certos 
par\^ametros t\^em os valores observados \'e um problema fundamental 
apenas {\it se} eles estiverem relacionados com objetos  
fundamentais. Talvez os quarks \nao sejam os objetos fundamentais da 
natureza. Por exemplo, os chamados \^angulos de Cabibbo-Kobayashi-Maskawa 
\sao equivalentes \`a orienta\cao de certas \'orbitas planet\'arias.
Esta orienta\cao \'e de fundamental import\^ancia para n\'os: 
ela determina as esta\coes na Terra. No entanto, \nao consideramos como 
fundamental explicar por primeiros princ\ii pios as orienta\coes das 
\'orbitas porque eles (os planetas) h\'a muito tempo deixaram de ser 
considerados objetos de estudo das 
leis fundamentais. N\ao era este o caso na \'epoca de, digamos, Kepler 
(ver mais adiante). 
Por enquanto consideramos os quarks como sendo fundamentais. 
Ser\'a isso mantido com o desenvolvimento da f\ii sica nos pr\'oximos 
d\'ecadas ou s\'eculos? N\ao sabemos. 

A ``inevitabilidade'' \'e outra carater\ii stica que Weinberg 
atribue \`a beleza de uma teoria~\cite{sw2}. A teoria da relatividade 
geral {\it parece} inevit\'avel uma vez adotados os princ\ii pios 
(simples) de Einstein. No entanto Weinberg subestima a inevitabilidade 
dos dados 
experimentais. Os dados astron\^omicos tornaram inevit\'avel a lei do 
inverso do quadrado da dist\^ancia. Nas outras intera\coes a 
inevitabilidade \'e obtida dando prioridade \`as simetrias em vez de a 
mat\'eria.

Um terceiro aspecto para Weinberg que deve ser incorporado \`a
beleza \'e a sua ``rigidez''~\cite{sw2}. 
Pode-se descrever uma grande variedade de fen\^omenos construindo-se 
teorias o mais flex\ii veis poss\ii veis. 
N\ao \'e isto o que esperamos de uma teoria dita de fundamental. 
A rigidez das teorias em \fpe \'e dada pela simetria e pela consist\^encia 
matem\'atica como por exemplo renormalizabilidade e o cancelamento das 
anomalias.


\section{Ca\c ca ao universo}
\label{sec:caca}
A procura da ``f\'ormula do mundo'' implica uma defini\cao do mundo.
Isto \'e, precissamos {\em a priori} definir o sujeito a ser explicado.
H\'a apenas alguns s\'eculos, o ``mundo'' era restrito aos planetas. Ainda 
que hoje em dia nosso ``mundo'' \'e mais complexo e amplo, 
\nao vemos nenhuma raz\ao pela qual j\'a tenham sido observados todas as 
suas carater\ii sticas a serem explicadas. Surpressas podem aparecer mesmo 
naquelas escalas espa\c co-temporais nas quais atualmente pensamos ja ter 
estudado em detalhe.

A postura adotada freq\"uentemente pelos f\ii sicos, \'e refletida
na vis\ao de Dirac. Segundo ele, a mec\^anica qu\^antica estava
completa em 1929 e as imperfei\coes relativas \`a sua s\ii ntese com a
relatividade restrita eram~\cite{sss}   
\begin{quotation}
{\em ...of no importance in the consideration of atomic and
mole\-cu\-lar 
structure and ordinary chemical reactions...the underlaying physical
laws necessary for the mathematical theory of a large part of physics
and the whole of chemistry are thus completely known, and the
difficulty is only that the exact application of these laws lead to
equations much too complicated to be soluble.}
\end{quotation}
Estas palavras de Dirac foram motivadas pelo sucesso da mec\^anica 
qu\^antica \nao relativ\ii stica na explica\cao da estrutura do \'atomo 
e mol\'eculas.

A vis\ao de Dirac \'e atualmente compartilhada pela maioria dos 
f\ii sicos. De fato, como deixa claro acima Schweber, o reducionismos \'e
a marca da f\ii sica te\'orica deste s\'eculo. Mais ainda, \'e uma
carater\ii stica, at\'e recentemente do\-mi\-nan\-te, de toda a ci\^encia 
moderna. N\ao \'e poss\ii vel negar os bons resultados obtidos. Ainda 
segundo Schweber~\cite{sss}
\begin{quotation}
{\em These conceptual developments in fundamental physics have revealed
a hierarchical structures of the physical words. Each layer of the 
hierarchy successfully represented while remaining largely decoupled
from other layers. These advanced have supported the notion of the 
existence of objective emergent properties and have challenged the 
reductionist 
approach. They have also given credence to the notion that to a high 
degree of accuracy our theoretical understanding of some domains have 
stabilized, 
since the foundational aspects are considered known.

Quantum mechanics reasserted that the physical world present itself 
hierarchically. The world was not carved up into terrestial, planetary and 
celestial spheres but layered by virtue of certain constants of nature...
Planck's constant allow us to parse the world into microscopic and 
macroscopic realms, or more precisely into the atomic and molecular domains 
and the macroscopic domains composed of atoms and molecules. The story 
repeated itself with the carring out of the nucleon domain: quasistable 
entities--neutrons and protons--could be regarded as the building blocks of 
nuclei, and phenomenological theories could account for many properties and 
interactions of nuclei. }
\end{quotation}

Os f\ii sicos te\'oricos \sao as vezes otimistas demais com rela\cao aos 
objetivos a serem alcan\c cados a curto pra\c co. 
Por\'em os f\'\i sicos tamb\'em s\ao c\'eticos, Weisskopf por exemplo se 
pergunta~\cite{vw1}
\begin{quotation}
{\em Is it really an aim of theoretical physics to get the world formula?
The greatest physicists have always thought that there was one, and that
everything else could be derived from it. Einstein believed it, Heisenberg 
believed it, I am not such a great physicist, I do not beleive it...
This I think, is beacuse nature is inexhaustible.}
\end{quotation}
Devemos perguntar-nos se o desenvolvimento futuro implica uma continua\cao
nessa dire\cao ou uma pausa para reorganizar todos os conhecimentos 
adquiridos at\'e hoje, antes de ser poss\ii vel a proposta de 
uma nova ordem. 

Por outro lado, acreditamos que o problema n\~ao \'e se devemos ou \nao 
reconhecer se a \fpe \'e a \'unica \'area fundamental 
da f\ii sica. O que estamos tratando \'e mais profundo. \'E se existem
leis verdaderamente fundamentais a serem descobertas (ou que j\'a
o tenham sido) em estruturas diferentes daquelas das pequenas escalas 
sub-nucleares ou no universo primordial.
\'E curioso observar que esse tipo de estruturas hier\'arquicas 
na dimen\sao espa\c co-temporal foram obtidas
sempre que os instrumento de observa\cao eram refinados para poder
atingir dist\^ancias cada vez menores. Por exemplo, o processo de dete\cao
e estudo de partes cada vez menores ocorre tamb\'em na biologia. Depois
de estudar doen\c cas bacterianas, com o advento do microsc\'opio
eletr\^onico, foram detetadas doen\c cas virais. Podem existir
agentes produtores de doen\c cas menores ({\sl prions}) ainda \nao 
detetados?~\cite{gallo}. \'E poss\ii vel que, al\'em de refinamentos na 
sensitividade dos aparelhos, que sem d\'uvida foi o eixo do desenvolvimento 
das ci\^encias, o refinamento das capacidades de c\'alculo possa introduzir 
novos conceitos. O caos pode ter sido um dos primeiros exemplos. A f\ii sica 
poderia entrar numa fase n\~ao reducionista (poderiamos dizer {\sl holista} 
mas este termo j\'a \'e usado com outros prop\'ositos; ou {\sl global} ou 
usar tamb\'em {\sl n\~ao-reducionista}).
Em todo caso pode ser que \nao seja uma reviravolta
completa. Os aspectos globais tem suas dificuldades tamb\'em e seu 
progresso \nao dever\'a ser t\ao r\'apido como alguns podem pensar. 
(Mencionamos antes 
que mesmo \'areas como o caos passam pelas mesmas dificuldades.) Por outro
lado, a tradi\cao reducionista ainda \nao foi esgotada e dever\'a dar 
resultados importantes nas pr\'oximas d\'ecadas. Segundo 
Weinberg~\cite{sw2}
\begin{quotation}
{\em  At this moment in the history of science it appears that the best
way to approach these laws is through the physics of elementary particle,
but is an incidental aspect of reductionism and may change.}
\end{quotation}

Argumento te\'oricos falharam as vezes redondamente. Vejamos por exemplo 
os seguintes argumentos de Maxwell~\cite{jcm}
\begin{quotation}
{\em ...to explain electromagnetic phenomena by means of mechanical action 
transmitted from one body to another by means of a medium occupying the 
space between them. The ondulatory theory of light also assume the 
existence of a medium.

To fill all space with a new medium whenever any new phenomena is to be 
explained is by no means philosophical, but if the study of two different 
branches of science has independently suggested the idea of a medium, and 
if the properties which must be attibutted to the medium...are the 
same...the evidence for the physical existence of the medium will be 
considerably strengthened.} 
\end{quotation}

O que Maxwell \nao sabia era 
que a estrutura matem\'atica da teoria dispensava a 
exist\^encia de qualquer meio para a transmis\sao de ondas 
eletromagn\'eticas. 
Por outro lado, historicamente a exist\^encia do medio para os fen\^omenos 
eletromagn\'eticos foi importante. N\ao apenas para Maxwell. Faraday, 
consideraba o v\'acuo como uma subst\^ancia. Isso ajudava-o a ver o campo 
eletromagn\'etico como sendo transmitido pelo meio. Isto foi um avan\c co 
com rela\cao \`a a\c c\~ao a dist\^ancia de Newton. 

A \fpe tem sido mesmo reducionista. E \'e a isso que deve seu sucesso. 
O ponto \'e se {\sl deve} continuar sendo, ou se chegou o momento de dar
\^enfase aos aspectos globais ou n\~ao-reducionistas. 
Assim colocada, esta discus\sao deixa de ser algo vazio. Ela pode determinar
o sucesso ou o fracasso de novas gera\coes de pesquisadores. Como foi dito
acima, \'e pos\ii vel que nas pr\'oximas d\'ecadas a tend\^encia na \fpe 
seja a mesma que a dos \'ultimos 50 anos. Tem muitos dados a serem obtidos 
antes de acharmos que devemos voltar a problemas mais fundamentais deixados 
para tr\'as (se \'e que isso acontecer\'a algun dia).

Por outro lado, devemos ter sempre em mente, a historicidade dos 
pro\-ble\-mas e de suas solu\c c\~oes. Na metade do Sec. XIX discutia-se se 
a cria\cao
espont\^anea da vida era poss\ii vel. Poderia a vida ter surgido da 
n\~ao-vida? As experi\^encias de Pasteur mostraram que o fen\^omeno de 
putrefa\cao era provocado pelos micro-organismos presentes no ar. A biologia 
era assim, separada da qu\ii mica. Esta separa\cao foi positiva nas d\'ecadas 
posteriores com ambas disciplinas se desenvolvendo separadamente. Mas, 
depois da mec\^anica qu\^antica passou-se a acreditar que todos os 
processos biol\'ogicos \sao reduzidos a processos qu\ii micos que pela 
sua vez \sao manifesta\coes das leis da f\ii sica elementar. Por\'em, 
em alg\'un 
momento da evolu\cao do universo (ou da Terra) a vida surgiu da n\~ao-vida 
num processo ainda n\ao conhecido. Apenas n\ao s\ao os processos simples
do dia-a-dia nos quais acreditavam os defensores da gera\cao espont\^anea
pre-Pasteur, por exemplo pela fermenta\c c\~ao  e putrefa\c c\~ao como 
acreditava F. A. Pouchet. 
Um outro exemplo, \'e a lei da gravita\c c\~ao de Newton. Como vimos na 
Sec.~2, os f\ii sicos europeos (Descartes principalmente), e o pr\'oprio 
Newton, \nao aceitavam do 
conceito de ``a\cao a dist\^ancia'' e do de ``espa\c co absoluto''. Mas, a 
lei de Newton da gravita\cao foi superior que a dos v\'ortices de Descartes 
para preparar o caminho da teoria da gravita\cao geral de Einstein.
Poderiamos colocar varios exemplos onde fica claro que uma solu\cao a um 
determinado problema permite o desenvolvimento de uma \'area mesmo que 
posteriormente se verifique que aquela solu\cao \nao era correta ou apenas 
o era de maneira aproximada.
O objetivo da ci\^encia continua a ser a {\it ca\c ca ao universo}. A 
discuss\ao \'e qual o passo mais imediato a ser dado na dire\cao certa.

\section{Part\ii culas  elementares: al\'em do modelo \\ padr\~ao}
\label{sec:mp}
Na d\'ecada dos anos 70 na \'area de \fpe ficou completo (do ponto de vista 
te\'orico) o chamado {\em modelo padr\~ao} no qual o mundo subnuclear \'e 
composto em termos de gluons, b\'osons vetoriais in\-ter\-me\-di\-a\-rios, 
o f\'oton, quarks e leptons e o escorregadi\c co b\'oson de 
Higgs~\cite{sm,fw,higgs}. Depois disso podemos  
perguntar-mos se haver\'a uma outra camada de estrutura. Como mencionado 
acima, n\ao sabemos. \'E por isso que a procura continua.
             
As id\'eias te\'oricas que permitiram \`a f\ii sica chegar ao 
estabelecimento de uma s\'erie de dom\ii nios hier\'arquicos quase 
aut\'onomos s\~ao: o grupo de renormaliza\cao (que nos indica como podemos 
fazer extrapola\c c\~oes), o teorema de desacoplamento (que nos permite 
esquecer ao fazer as extrapola\c c\~oes, part\'\i culas de massa maior 
que a escala de energia relevante para as experi\^encias), a liberdade 
assint\^otica (que nos 
permite usar teoria de perturba\c c\~oes) e a quebra esp\^ontanea de 
simetria (que nos permite gerar massa para as diferrentes part\'\i culas 
sem estragar a consist\^encia matem\'atica da teoria). 

O sucesso deste modelo na descri\cao das intera\coes entre part\ii culas 
e\-le\-men\-ta\-res coloca o problema de se determinar quais as leis da 
f\ii sica al\'em deste modelo. A despeito da impressionante concord\^ancia 
com os dados experimentais, existe um concenso entre os f\'\i sicos de que 
este modelo \nao \'e a teoria final. 
O modelo deixa muitas coisas sem resposta e tem muitos par\^ametros a serem 
determinados pela experi\^encia. Como mencionamos antes, isso poderia \nao 
ser um problema j\'a que qualquer teoria f\ii sica vai precissar sempre de 
um n\'umero (finito) de par\^ametros de entrada  a serem determinados pela 
experi\^encia. 

O ponto de vista reducionista implica 
que tudo na natureza \'e controlado por um mesmo conjunto de leis
fundamentais. O modelo padr\ao estaria na base de tudo o resto mas e 
depois, o que \'e que explica esse modelo? quais os princ\ii pios gerais 
que explicariam 
porque esse modelo e \nao outro \'e o que \'e v\'alido at\'e as energias dos 
aceleradores atuais. Constitue este um problema fundamental a servir de guia 
para as futuras gera\c c\~oes? A resposta usual a esta pergunta est\'a no 
esp\ii ritu das palavras de Einstein quem, em 1918, dizia  
\begin{quotation}
{\em The supreme test of the physicist is to 
arrive at those universal elementary 
laws from which the cosmos can be buildt by pure deduction}.
\end{quotation}
Tarefa dif\ii cil, nem mesmo sabemos como construir {\it por primeiros 
princ\ii pios} hadrons partindo de quarks e gluons! (Para \nao falar de 
n\'ucleos em termos de nucleons, mol\'eculas
em termos de \'atomos.) Esse tipo de afirma\cao emotiva, mesmo vindo de
f\ii sicos como Einstein devem ser analisadas cuidadosamente. 
Principalmente pelos estudantes que est\ao come\c cando a sua 
p\'os-gradua\c c\~ao. 

Atualmente a \'area de neutrino \'e uma das mais ativas da \fpe fornecendo 
muitos dados experimentais que permitem testar hip\'oteses do que seria a 
f\ii sica {\it al\'em} do modelo padr\~ao. De maneira geral as 
observa\c c\~oes  astrof\ii sicas~\cite{gefan} se 
unem aos dados de aceleradores e de experimentos com energias baixas que 
est\ao medindo com maior preciss\ao efeitos bem conhecidos para a procura
da nova f\'\i sica. Novos dados de efeitos h\'a muito tempo procurados
como por exemplo a 
viola\cao da simetria CP~\cite{fermilab,cern}, ou novas possibilidades 
permitidas por novas t\'ecnicas experimentais como o estudo da 
anti-mat\'eria com a produ\cao e armazenamento de anti-pr\'otons e mesmo
de anti-hidrog\^enio~\cite{cern}. Onde est\'a a crise? 

\section{Dire\c c\~ao \'unica?}
\label{sec:edai}

Por ser reducionista a ci\^encia moderna \'e tamb\'em unificadora. 
Unificadora no sentido que pretende uma descri\cao unificada 
dos fen\^omenos f\ii sicos e reducionista no sentido que pretende
reducir o n\'umero de conceitos independentes com os quais seriam
formuladas as leis da natureza. Esse ponto de vista
foi criticado por Anderson alguns anos atr\'as. Segundo ele~\cite{pwa}
\begin{quotation}
{\em The main fallacy in this kind of thinking is that the reductionist
hypothesis does not by any means imply the ``constructionist'' one:
The ability to reduce everything to simple fundamental laws does not
imply the ability to start from those laws and reconstruct the
universe. In fact, the more the elementary particle physics tell us
about the nature of fundamental laws, the less relevance they seem to
have to the very real problems of the rest of science, much less to
those of society. 

The constructionist hypothesis breaks down when confronted with the
twin difficulties of scale and complexity. The behavior of large and
complex aggregates of elementary particles, it turns out, is not to
be understood in terms of a simple extrapolation of the properties of
a few particles. Instead, at each level of complexity enterily new
properties appear, and the understanding of the new behaviors
requieres research which I think is as fundamental in its nature as
any other.}
\end{quotation}

O quadro do percurso desde o ``menos fundamental'' at\'e o ``mais
fundamental'' pode ser resumido na Tabela 1 na
qual os elementos de uma ci\^encia $X$ obedecem as leis de uma
ci\^encia $Y$~\cite{pwa}.
\vskip .5cm
\begin{table}
\begin{tabular}{|c||c|}\hline
X & Y \\ \hline
Estado s\'olido ou Muitos Corpos & Part\ii culas Elementares \\
Qu\ii mica & Muitos corpos \\
Biologia Molecular &  Qu\ii mica \\
Biologia Celular & Biologia Molecular \\
$\vdots$ & $\vdots$ \\
Psicologia & Fisiologia \\
Ci\^encias Sociais & Psicologia \\ \hline
\end{tabular}
\caption{``Hierarquia'' das ci\^encias de Anderson.}    
\end{table}                   
\vskip .8cm
A hierarquia mostrada na Tabela 1, por\'em, \nao implica que a              
ci\^encia X seja {\it apenas} a\-pli\-ca\-\cao da ci\^encia Y. 
Em cada n\ii vel novas leis, conceitos, generaliza\coes e mesmo novos 
m\'etodos de pesquisa \sao necess\'arios. 
Mesmo que saibamos que ap\'os o aquescimento as
mol\'eculas se afastam at\'e que a forma s\'olida se dissocie, as
mol\'eculas agora obedecem as leis dos fluidos que \nao podem ser
deduzidas a partir das leis dos s\'olidos. Por exemplo, a vida ( a 
biologia em geral) \'e em seu n\ii vel mais fundamental, qu\ii mica. 
Isso \nao implica
que seja {\em apenas} qu\ii mica. O mesmo pode ser dito da qu\ii mica,
ela \'e basicamente f\ii sica mas as leis da f\ii sica ajudam pouco no
estabelecimento de novas leis qu\ii micas. Claro que essas novas
leis da qu\ii mica n\ao devem violar as leis da f\ii sica. 
Mas, fazer qu\ii mica \nao \'e fazer f\ii sica. E nunca ser\'a.

Na pr\'atica temos ``disconnected clumps'' nos diferentes dom\ii nios das 
ci\^encias. Isso acontece nos dois sentido referidos acima: um dom\ii nio 
de sub-estrutura \nao ajuda na explica\cao da maioria dos processos da 
estrutura acima dela.  Para refor\c car o assunto enfatissemos que at\'e 
pouco tempo 
atr\'as a  f\ii sica at\^omica entra como um fator de corre\cao da f\ii sica 
nuclear. Esta pela sua vez \nao \'e ``construida'' (no sentido de Anderson) 
pela f\ii sica de quarks. Mas acreditava-se que por sua vez a f\'\i sica
nuclear n\~ao teria nada a ver com a f\ii sica at\^omica. 
No entanto, recentemente foi descoberto um efeito que contradiz esta 
\'ultima afirma\c c\~ao: foi encontrado que a orienta\c c\~ao do spin 
do n\'ucleo de uma mol\'ecula de $H_2$ afeta o espalhamento dessa 
mol\'ecula biat\^omica na superficie de um cristal~\cite{bertino}. Isso vai
ajudar a estudar estrutura do campo el\'etrico em superf\ii cies.
At\'e onde isso pode ir? isto \'e, ser\'a que um dia estaremos observando
efeitos do conte\'udo de quarks em f\'\i sica do estado s\'olido?
n\~ao sabemos, isso depende de melhoramentos na t\'ecnica que est\~ao fora
da nossa capacidade de previs\~ao. Mas, se isso acontecer ent\~ao Anderson 
estaria errado!

\'E usual acreditar que quando encontradas, verdades universais devem 
ser explicadas em termos de outras mais profundas,..., at\'e atingirmos 
a chamada {\sl teoria final}. Este \'e de fato {\sl um} dos projetos para 
a ci\^encia. 
Mas \nao \'e o \'unico. E nem mesmo talvez seja o mais interessante.
Um princ\ii pio cient\ii fico explica outro se este \'ultimo \nao viola
as leis do primeiro. Por\'em, temos de entender que as leis do 
princ\ii pio mais b\'asico \nao ajudam a determinar as leis do segundo. 
Apenas servem
como referencial subjacente. \'E por isso que continuar\'a havendo 
qu\ii mica independentemente de que de seus fundamentos sejam 
f\'\i sicos. Mais ainda, 
as leis da qu\ii mica ou da mat\'eria condensada, para p\^or dois exemplos, 
podem ter uma generalidade vertical (no sentido de Weisskopf acima).
Podemos colocar a evolu\cao do progresso cient\ii fico da 
forma mostrada na Tabela 2~\cite{dresden}

\begin{center}
\begin{table}[ht]
\begin{tabular}{|l||l||l|}\hline
&1) observa\c c\~oes, fen\^omenos  &\\
&complexos, infinidade & \\
$\downarrow$ &  de objetos, & $\uparrow$ \\
  & 2) Organiza\c c\~ao em termos  & Introdu\c c\~ao \\
& de conceitos emp\'\i ricos & de detalhes \\ 
Redu\c c\~oes &    & conceitos 
``\'uteis''\\
sucessivas  &3) Leis emp\'\i ricas--no\c c\~ao de  & \\
 & objetos compostos &  \\
$\downarrow$ & 4) As leis emp\'\i ricas  & $\uparrow $ \\
             &   podem ser expressas &\\
             &   como rela\c c\~oes formais  & 
especializa\c c\~ao  \\          
  &  5) Poucos objetos simples, &  \\
 &  leis mais gerais &  \\
&  6) Abstra\c c\~oes, matematiza\c c\~ao,  & \\
&  idealiza\c c\~ao, generaliza\c c\~ao  &  \\
&  7) Objetos simples irredut\'\i veis,  & $\uparrow$ \\
 &  conceitos e rela\c c\~oes  & O problema  \\
O fim&  universais, & inverso:\\
a redu\c c\~ao & leis   & da redu\c c\~ao \\
completa &   &  \`a composi\c c\~ao \\\hline
\end{tabular}
\caption{Seq\"u\^encias do reducionismo vs composi\c c\~ao.}
\end{table}
\end{center}


A vis\ao de Dirac continua na tradi\cao da f\ii sica de part\ii culas
e campos. Na d\'ecada dos anos 80 as teorias de supercordas que tinham sido
elaboradas desde 1974 por Veneziano, pasando por Nambu e outros, 
apareceram como fortes candidatas para a teoria que unificasse as quatro
intera\coes conhecidas. Essa seria ent\ao a culmina\cao da vis\ao
reducionista da f\ii sica.
Segundo Witten~\cite{do} a teoria das supercordas s\~ao  
\begin{quotation}
{\em ...a piece of twenty-first-century physics that has
fallen into the twentiesth century, and would probably require
twenty-second-century mathematics to understand}.
\end{quotation}
Em 1980 Hawking disse~\cite{sh} que existia  
\begin{quotation}
{\em ...the possibility that the goal of the theoretical physics might be
achieved in the not too distant future, say, by the end of the
century. By this I mean that we might have a complete, consistent and
unified theory of the physical interactions which would describe all
possible observations.}
\end{quotation}
Isso tem mais de pessimista que de otimista. Significa que todo o 
conhecimento te\'orico e em particular novos dados experimentais \nao 
ser\ao capazes, nos pr\'oximos s\'eculos, de indicar uma outra dire\cao 
para as leis da natureza. Esse \'e o ponto fraco de todo o paradigma de 
``unifica\c c\~ao''. 

Essa posi\cao come\c ca a mudar. Um exemplo radical \'e o de Georgi quem 
afirma que
\begin{quotation}
{\em It {\it is} true that in chemistry and biology one does not encounter
any new physical principles. But the systems on which the old principles
act differ in such a way drastic and qualitative way in the different 
fields
that it is simply not {\it useful} to regard one as a branch of another.
Indeed the system are so different that `principles' of new kinds must be
developed, and it is the principles which are inherently chemical or 
biological which are important.

In the same way, to study phenomena at velocities much less than
$c$ and angular momentum much greater than $\hbar$, it is simply not 
useful to regard them as special cases of phenomena for arbitrary velocity
and angular momentum. We do not need relativity and quantum mechanics for
small velocity and large angular momenta...if we had to discover the laws
of relativistic quantum mechanics from the beginning, we probably
would never have gone anywhere.}
\end{quotation}

Nenhuma forma de estudar a natureza \'e compar\'avel \`a pesquisa cient\ii 
fica a partir (principalmente) de Galileo. Entendemos quantitativamente os 
fe\-n\^o\-me\-nos. Isso faz, entre outras coisas, a diferen\c ca entre 
nossos \'atomos e os de Dem\'ocrito. Entender quantitativamente os 
fen\^omenos diz 
respeito a que podemos fazer predi\coes quantitativas que podem ser 
confirmadas ou \nao pela experi\^encia.~\footnote{N\ao pretendemos que 
todos 
os aspectos de uma teoria tenham que ser testados pela experi\^encia. 
Esta era a posi\cao dos positivistas. 
As teorias segundo eles t\^em de estar baseadas 
{\em apenas} em observ\'aveis.} Sem estas \'ultimas \nao podemos dizer se 
uma teoria \'e correta ou n\~ao. Claro, as coisas n\ao s\ao t\ao simples 
como parecem dado que podem existir segundo os dados
experimentais v\'arias teorias poss\ii veis. Aqui a simplicidade \'e 
\'util. Mas apenas isso, \'util, n\ao definitiva. 
Assim, a divi\sao aristot\'elica de movimentos 
{\sl naturais} e {\sl \nao naturais} \nao passa de uma descri\cao cuja 
plausabilidade \nao pode ser testada. Mesmo que a f\ii sica mo\-der\-na 
fizesse uso de tais conceitos (como o faz do \'atomo) devemos distinguir 
uma opini\ao de uma pesquisa metodol\'ogica (mais ou menos) bem definida.
Neste sentido, a refer\^encia aos \'atomos de Dem\'ocrito \'e apenas
aned\'otica.

\'E bom lembrar que ainda que f\ii sicos como Newton e Faraday tinham em 
mente uma ``teoria final'', agora sabemos que o contexto te\'orico e 
experimental da \'epoca era bem restrito para tal efeito. 
Isso \'e mais um exemplo de que o 
m\'etodo cient\ii fico (qualquer coisa que isso signifique) \nao \'e 
suficiente para explicar as motiva\c c\~oes, as escolhas e os preconceitos 
dos cientistas. Assim, acreditamos que a quantidade de especula\cao \'e 
restrita por fatos al\'em das opini\~oes da comunidade cient\ii fica e 
{\it a priori} \nao est\'a bem definida.   

Os exemplos da teoria geral da relatividade e da predi\cao da radia\cao 
de fundo \sao exemplos de extrapola\coes que deram certo e isso motivou a 
extrapola\cao dos resultados te\'oricos al\'em das possibilidades de 
verifica\cao experimental. Mas, quantas extrapola\coes falharam? 
No m\ii nimo para  sermos consistentes com a estat\ii stica devemos 
considerar isso quando fizermos es\-co\-lhas pessoais sobre o tema de 
pesquisa. 
O caso contr\'ario tamb\'em acontece. Achar que tudo j\'a \'e conhecido 
e que \nao h\'a mais espa\c co para especula\c c\~oes. \'E bem conhecida 
a opini\ao no final do s\'eculo pasado (atribu\ii da a Lord Kelvin) e 
mesmo no come\c co deste s\'eculo (como Michelson) sobre o 
fato que tudo que tinha de ser descoberto j\'a o tinha sido feito.
Assim, podemos nos perguntar se a luta de Einstein \'e ainda a nossa. 
N\ao no sentido escatol\'ogico no qual \nao temos a menor d\'uvida que 
\'e. Mas no sentido de uma escolha pessoal da linha de pesquisa de um(uma) 
jovem cientista. 

Que existe um sentido nas explica\coes \nao h\'a d\'uvida: as leis de 
Newtonexplicam as de Kepler, as de Einstein as de Newton, etc. O ponto 
\'e, se esse
sentido \'e \'unico ou, existem ramifica\c c\~oes? Quando uma teoria final
no sentido 
$$\cdots\rightarrow\mbox{mol\'eculas}
\rightarrow\mbox{\'atomos}\rightarrow 
\mbox{n\'ucleos}\rightarrow\mbox{n\'ucleons} \rightarrow \mbox{quarks} 
\rightarrow\cdots$$
for obtida, ainda fen\^omenos como a turbul\^encia e supercondutividade a
altas temperaturas precissar\ao ser 
explicados e o que esteja para al\'em dos quarks poder\'a \nao ser 
importante para isso.

Se podemos dizer que as verdades mais fundamentais \sao aquelas mais
abrangentes devemos, no entanto, aceitar que existem verdades fundamentais
``horizontais'' (``extensivas''  no sentido de Weisskopf), isto \'e, \nao 
fazem parte de uma mesma cadeia de explica\coes em ordem crescente da 
escala de determinadas grandezas (massa, velocidade ou 
energia). Assim, as leis de Newton podem ser mais fundamentais que as de 
Kepler e as de Einstein, pela sua vez, mais fundamentais que as de Newton. 
Mas, ser\'a que isso ajudaria na compreen\sao das propriedades do ADN? 
ser\'a que apenas seria necess\'ario um grande computador para explicar 
essas propriedades resolvendo equa\coes da mec\^anica qu\^antica para os 
el\'etrons e os n\'ucleos?  Talvez n\~ao. Do contr\'ario teriamos voltado 
ao mecanicismo pre-Maxwell, apenas substituindo a mec\^anica cl\'assica 
pela qu\^antica. Podem existir quest\~oes que \nao possam ser resolvidas 
com as nossas ferramentas atuais, te\'oricas ou experimentais. 

Alguns fatos, como a origem da vida, parecem ser devidos a acidentes 
hist\'oricos.
Se, contudo, alg\'um dia as condi\coes iniciais passassem a ser parte das
leis da f\ii sica  isso pode ser feito n\~ao necessariamente no sentido
$\mbox{\'atomo}\rightarrow \mbox{n\'ucleon}\cdots$ mas, mesmo com 
fen\^omenos macrosc\'opicos. Novos princ\ii pios que \nao contradigam as 
leis microsc\'opicas poder\ao encontrar novas generalidades \nao 
deduz\ii veis daqueles. 

Por exemplo, a universalidade do caos \'e suficientemente abrangente
e \nao depende (por enquanto) de leis mais gerais em escalas menores.
\'E este tipo de universalidade que acreditamos existir em diferentes 
n\ii veis de organiza\cao independentes uns dos outros.
Por outro lado, atualmente existem teorias t\ao especulativas 
(a teoria dos ``baby universes'' e outras) \nao completamente formuladas 
matematicamente e sem suporte experimental (mesmo a longo prazo) que 
podemos at\'e compar\'a-las com a formula\cao aristot\'elica (a 
situa\c c\~ao {\it grega} mencionada antes).

N\ao \'e obvio que vai acontecer com o caos o que aconteceu com a 
termodin\^amica. Esta come\c cou como ci\^encia aut\^onoma mas foi depois 
fundamentada na me\-c\^a\-ni\-ca estat\ii stica. Muito menos obvio \'e  
ocaso da biologia ou do problema da conci\^encia~\cite{jh2,rp}.
Ainda que a mec\'anica estat\ii stica ``explica'' a termodin\^amica apenas
no sentido que a incorpora.~\footnote{As cr\ii ticas a Boltzmann estavam
corretas porque apenas a mec\^anica qu\^antica permitiria uma formula\cao
coerente das leis   estat\ii sticas mas ela n\ao era conhecida nos
primeiros anos do s\'eculo~\cite{tk2}.}

Mas devemos ser cr\ii ticos tamb\'em com rela\cao a essas posi\c c\~oes.
\'E verdade que \nao adianta muito para os qu\ii micos saber que a 
materia \'e formada por quarks. Mas de alguma maneira esse conhecimento
\'e subjacente a toda a qu\ii mica. Gostemos ou n\~ao. Na pr\'atica nossos 
m\'etodos te\'oricos \sao muito limitados. \'E sempre
dif\ii cil considerar as situa\coes limites como aquele entre a mec\^anica
qu\^antica e a cl\'assica, ou como diz Georgi acima, entre a mec\^anica 
relativ\ii sta e a n\~ao-relativ\ii sta. Mas essas dificuldades devem ser
vistas como limita\coes nossas e \nao \sao \'unicas nessas \'areas.

Acontecem mesmo na mec\^anica cl\'assica n\~ao-relativ\ii sta.
Por exemplo, sabemos que as diferentes maneiras de formular a mec\^anica
cl\'assica como 1) leis de Newton, 2) princ\ii pio de D'Alembert,
3) princ\ii pio dos deslocamentos virtuais, 4) princ\ii pio de Gauss,
5) princ\ii pio de Hamilton, 6) princ\ii pio de a\cao m\ii nima,
7) coordenadas generalizadas e equa\coes de Lagrange, 8) equa\coes
can\^onicas de Hamilton, 9) equa\coes de Hamilton-Jacobi e teoria das
trasforma\c c\~oes. Todos estes formalismos s\~ao completamente 
equivalentes no sentido que, qualquer problema de mec\^anica cl\'assica
pode, em princ\ii pio, ser resolvido por qualquer um desses m\'etodos. 
(Na pr\'atica porque todos levam \`as equa\coes de Newton.)
As vezes, alguns deles \sao mais ou menos apropriados para um problema 
particular. Outro, t\^em a vantagem de permitir uma aprecia\cao mais 
profunda dos sistemas din\^amicos. Finalmente, alguns deles \sao mais 
apropriados na respectiva exten\sao qu\^antica~\cite{longair}. 
Contudo, nem toda 
informa\cao \'e a mesma em cada um destes formalismos. Por exemplo, com 
rela\cao as simetrias e leis de conserva\c c\~ao. A conserva\cao da energia, 
momento linear e momento angular aparecem em qualquer dos formalismos acima 
mencionados. Mas, em geral as leis de conserva\cao podem ser diferentes. 
As simetrias do sistema \sao diferentes quando se usam as equa\coes do 
movimento ou a Lagrangeana. Qu\'al deles seria mais fundamental? Lembremos 
que algumas equa\coes do movimento nao t\^em uma Lagrangeana ou 
Hamiltoniana correspendente. N\ao existem respostas definitivas dentro 
dos nossos conceitos te\'oricos atuais para esse tipo de pergunta. Nem 
por isso achamos que eles \nao descrevem a mesma mec\^anica cl\'assica.

Por outro lado, devemos lembrar que algumas ci\^encias s\~ao por natureza
pr\'opria {\sl globais} por exemplo as chamadas {\it Ci\^encias da 
Terra}~\cite{brown}.
Nesta nova s\ii ntese a Terra \'e considerada como sendo um {\it sistema} 
cuja din\^amica reg\^e-se por causas m\'ultiplas
que se ligam e regulam entre si~\cite{allegre}. A moral da hist\'oria \'e 
que Terra n\~ao pode ser tratada de jeito nenhum de maneira reducionista.
Constitue um problema suficiente geral.~\footnote{As suas leis poder\ao ser 
verificadas em planetas diferentes da Terra quando forem estudados. 
Recentemente foram encontradas evid\^encias de que no planeta Marte houve
invers\~oes do campo magn\'etico o que implicaria uma tect\^onica de 
placas semelhante \`a da Terra~\cite{marte}}

\section{Caos: leis fundamentais?}
\label{sec:x}

A ferramenta em \fpe para a extrapola\cao das leis de uma
determinada escala para escalas menores \'e o {\em grupo de renormaliza\c
c\~ao}~\cite{kw}. Sabemos, ent\~ao, como extrapolar leis conhecidas a uma 
determinada escala de dist\^ancias para escalas menores. Por\'em, se novas 
leis ser\~ao des\-co\-ber\-tas no futuro, e \nao vemos 
nenhum princ\ii pio geral que o proiba, ent\ao deveremos ir atualizando 
nossas ex\-tra\-po\-la\c c\~oes. Assim, qualquer afirma\cao relativa ao 
futuro do universo como um todo deve ser entendida apenas como uma 
predi\cao dos nossos conhecimentos {\em atuais} das leis fundamentais. 
Essas afirma\coes mudar\ao quando novas leis fundamentais sejam
descobertas nas escalas intermedi\'arias ou mesmo na dire\cao horizontal. 
Quem poderia 
ter previsto a descoberta da radioatividade? ou a mec\^anica qu\^antica 
poderia ter sido postulada apenas por m\'etodos formais? Al\'em disso, tudo 
est\'a baseado numa hip\'otese que mesmo razo\'avel poderia \nao ser 
verdadeira: a que as leis da natureza foram sempre as mesmas. Claro, \nao
existe uma proposta razo\'avel para uma poss\ii vel
varia\cao temporal dessas leis. A proposta de Dirac, que as constantes 
da natureza podem variar com o tempo n\ao foi confirmada at\'e agora e 
pode \nao ser a mais interessante~\cite{dirac}.

Devemos perceber que, se nem a arg\'ucia nem a estupidez
\sao previs\ii veis muito menos o \sao as futuras descobertas
te\'oricas e/ou experimentais. De qualquer forma, a Natureza \'e mais
imaginativa do que n\'os. A eleg\^ancia matem\'atica \nao \'e
suficiente. Podemos imaginar quais seriam as estruturas matem\'aticas
se os f\ii sicos do s\'eculo passado tivessem tentado unificar, mais ou
menos no sentido que conhecemos hoje, a eletrodin\^amica de Neumann e
Weber com a gravita\cao de Newton? Nessa eletrodin\^amica as
for\c cas eletromagn\'eticas se propagam de um corpo a outro com
velocidade infinita. Teriam resistido essas estruturas matem\'aticas
\`as descobertas experimentais do final do se\'culo XIX? \'E bem
prov\'avel que n\~ao. De fato, \'e interessante observar que Faraday 
queria mostrar que o eletromagnetismo estava relacionado com a 
gravita\c c\~ao~\cite{jh}. Isso mostra, repetimos, que as motiva\coes 
pessoais dos cientistas \nao tem nada a ver com os resultados reais 
obtidos. Faraday ficou longe de atingir seu desejo. Mas, visto 
restrospectivamente, ser\'a que precissava dele?  

Em 1950 John Von Neumann construia se computador Johnniac ({\em sic}).
A\-cre\-di\-ta\-va Von Neuman que a metereologia seria a \'area
principal do uso dos computadores~\cite{dyson}. 
Segundo Von Neumann os fen\^omenos metereol\'ogicos eram de dois
tipos: os {\sl est\'aveis} e os {\sl inst\'aveis}. Os primeiros s\ao 
aqueles que suportam pequenas perturba\c c\~oes, os segundos n\~ao. 
Por\'em, assim que os computadores estivessem funcionando todos os 
problemas relativos \`a predi\cao do tempo seriam resolvidos. Todos 
os processos est\'aveis seriam previstos e os inst\'aveis controlados. 

Von Neumann n\ao imaginou que \nao \'e poss\ii vel classificar a
desloca\cao de fluidos em previs\ii veis e control\'aveis. N\ao
previu a descoberta do {\em caos determin\ii stico}~\cite{ruelle}. 
Este fen\^omeno \'e caraterizado
por uma dep\^endencia hipersens\ii vel das condi\coes inicias,
quaisquer que sejam estas condi\c c\~oes. Isso quer dizer que neste
tipo de sistemas, pequenas perturba\coes implicam grandes efeitos a
longo prazo. 

O movimento regido pelas leis da mec\^anica newtoniana
\'e determinado sem ambig\"uidade pela condi\cao inicial, no entanto,
existe, em geral, uma limita\cao na predi\cao de sua trajet\'oria.
Temos ent\~ao, ao mesmo tempo determinismo e impreditibilidade a
longo prazo. O que define um sistema din\^amico \'e uma evolu\cao
temporal determinista bem definida. 
Talvez seja interessante observar que toda a f\ii sica desde os
gregos at\'e poucos anos atr\'as foi baseada na geometria cl\'assica
(euclideana ou n\~ao) na qual os elementos b\'asicos das formas 
\sao as linhas, planos, c\ii rculos, esferas, cones, etc. No entanto a
geometria {\em fractal}~\cite{fractal} parte de um universo mais
parecido ao real: irregular e \'aspero. Podemos nos perguntar quais
seriam as leis b\'asicas se este tipo de geometria fosse o
paradigma desde o come\c co. Ser\'a que o caos, seria um fato
incorporado nas pr\'oprias leis do movimento, em vez de s\^e-lo nas
condi\coes iniciais, como ocorre quando 
consideramos as leis de Newton? De qualquer forma o caos e a
geometria fractal da natureza \'e a vingan\c ca de Simplicio sobre
Sartori~\cite{galileo}. O movimento real n\~ao \'e t\~ao simpels como
acreditava Galileu. (Este \'e mais um exemplo de que a escolha de
teorias ou resultados tem um car\'ater hist\'orico. A teoria de Galileo
se mostrou frut\ii fera entanto que a vis\ao global n\~ao o foi. Mas 
acabariam se encontrando!).

Do ponto de vista conceitual a descoberta do caos \'e uma revolu\cao
como o foram as teorias da relatividade e a mec\^anica qu\^antica. No
entanto trata-se de fen\^omenos a grandes escalas, inclusive com 
rela\cao a escala humana. 
Assim, vemos que este poderia ser um exemplo de que as ``leis 
fundamentais'' aparecem \nao necessariamente quando estudamos
processos carater\ii sticos de dimens\~oes cada vez menores. 
Um aspecto a ser levado em conta \'e a ``universalidade'' de qualquer 
coisa que possamos chamar de ``lei fundamental''. 

A depend\^encia hipersens\ii vel das condi\coes iniciais foi
descoberta no final do s\'eculo XIX por Jacques Hadamard.
Contribui\coes importantes foram feitas por Duhem e 
Poincar\'e. No entanto apenas com o advento dos computadores
r\'apidos foi poss\ii vel fazer um estudo quantitativo riguroso.
Assim, podemos dizer que a coloca\cao do caos como um novo paradigma
\'e um feito que come\c cou na d\'ecada dos anos 60. Isso significa
que foram precissos mais de tr\^es s\'eculos para que novas
fen\^omenos com suas respectivas leis fossem descobertos
``dentro das leis fundamentais'' de
Newton. Neste sentido, poderiamos comparar as experi\^encias realizadas
a altas energias como equivalentes \`a experi\^encia de Cavendish:
apenas est\ao tentando descobrir generalidades sobre leis
fundamentais. O estudo detalhado fica como tarefa para as pr\'oximas
d\'ecadas (s\'eculos ?).   

Assim, voltando a von Neumann, ele \nao imaginou que em alguns anos
seria descoberto que o movimento ca\'otico que geralmente \'e
imprevis\ii vel e incontrol\'avel \'e que \'e a regra \nao a exce\c c\~ao. 
Vemos ent\~ao, e poderiamos dar muitos mais exemplos, que a preditividade
\'e pequena mesmo para mentes como as de Von Neumann.

\section{Que biologia \'e essa?}
\label{sec:biologia}
Nos dias de hoje \'e frequente escutar que ``assim como a f\ii sica foi a 
ci\^encia do s\'eculo XX a biologia ser\'a a ci\^encia do s\'eculo XXI''.
De fato, da d\'ecada de 50 para c\'a os avan\c cos na biologia molecular
s\~ao impressionantes. A\-tu\-al\-men\-te os projetos de sequ\^enciamento 
dos genomas de v\'arios organismos, em particular o projeto Genoma 
Humano~\cite{mt,science} permite visualizar um sim fim de aplica\c c\~oes 
da gen\^omica nas \'areas da sa\'ude e agropecuaria. 
At\'e tem sido dito que os f\ii sicos deveriam fazer biologia. 
Afortunadamente, quando a f\ii sica
estava realizando as suas hoje famosas descobertas nas tr\^es primeiras
d\'ecadas deste s\'eculo os biologos continuaram ... a fazer biologia! 
Um fato, no entanto, deve ser enfatizado. A biologia realizando essa
es\-pe\-ta\-cu\-lar revolu\c c\~ao fica menos biologia no sentido 
tradicional. A biologia est\'a se convertindo cada vez mais em uma 
ci\^encia quantitativa como a qu\ii mica e a f\ii sica. A matem\'atica 
e a inform\'atica s\~ao cada
vez mais impressind\ii veis para continuar o seu desenvolmimento. 
Sem os programas
sequenciadores n\~ao teria sido poss\ii vel realizar os projetos Genoma.
Craig Venter da Celera Genomics est\'a instalando o segundo maior 
conglomerado de computadores do mundo (somente inferior ao do Departamento 
de Energia
dos Estados Unidos)~\cite{mt}. Segundo Leroy Hood ``a biologia se tornou 
informa\c c\~ao'', metade dos cientistas que tra\-ba\-lha\-r\~ao no 
instituto que ele est\'a montando na Universidade de Seattle ser\ao 
matem\'aticos,
f\ii sicos, cientistas da computa\c c\~ao e qu\ii micos~\cite{mt}.
O genoma humano n\ao diz como os 100 mil genes trabalham juntos para 
formar o organismo humano. A compreens\ao disso \'e uma tarefa que n\ao 
pode ser levada adiante somente pelos biologos. Esse \'e um empreendimento 
multidisciplinario no qual os f\ii sicos poder\ao fazer contribui\coes
importantes. N\ao apenas eles, para entender o modo como as diferentes 
partes de qualquer genoma interagem entre si ser\ao necess\'arios n\ao 
apenas computadores cada vez mais r\'apidos e programas cada vez mais 
sofisticados,
o que da origem a uma nova \'area a {\it bioinform\'atica}, mas tamb\'em
ser\'a necess\'ario construir modelos matem\'aticos e estat\ii sticos,
compreender melhor a intera\cao entre as mol\'eculas, tarefa para 
qu\ii micos.
 
De fato a influ\^encia dos f\ii sicos em outras \'areas das ci\^encias fica 
evidente quando vemos que: M. F. Perutz, ganhou o pr\^emio Nobel de 
Qu\ii mica em 1962 pelos 
seus estudos da estrutura das proteinas globulares. No mesmo ano F. H. C. 
Crick ganhava o pr\^emio  Nobel de Fisiologia e Medicina pela descoberta 
da estrutura da dupla h\'elice do DNA. Em 1962 foi a vez de M. Delbr\"uck
pela descoberta do mecanismo de replica\cao e a estrutura gen\'etica dos
virus (em f\ii sica temos o espalhamento Delbr\"uck). 
 W. Gilbert ganhou em 1980 o pr\^emio Nobel de Qu\ii mica pelos estudos
na bioqu\ii mica dos \'acidos nucleicos em particular do DNA recombinante 
(em f\ii sica \'e conhecido por sua demostra\cao do teorema de Goldstone) e 
apenas para citar o mais recente, em 1998 o pr\^emio Nobel de qu\ii mica
teve um f\ii sico entre os ganhadore, W. Kohn pelas suas contribui\c c\~oes 
\`a qu\ii mica computacional.  
Sim, alguns f\ii sicos continuar\ao a fazer biologia mas a 
forma\c c\~ao tradicional de biol\'ogos (e m\'edicos e outras carreiras 
afins) ter\'a de ser reformulada.
  
\section{Computa\c c\~ao qu\^antica}
\label{sec:cq}

As contribui\coes dos f\ii sicos \`a \'area da computa\cao t\^em sido
tamb\'em impressionantes. E n\ao devemos esquecer que isso foi obtido sem 
ter como motiva\cao a aplica\cao que posteriormente apareceu. A World 
Wide Weg (WWW) foi desenvolvida no CERN (usando a j\'a 25 anos de velha 
Internet) com outras finalidades~\cite{jbi}. A revolu\cao somente 
aconteceu, no entanto, 
quando foi desenvolvido o Mosaic no NCSA (National Center for 
Supercomputer Applications) 

Todas as \'areas sem exe\c c\~ao t\^em sido influenciadas pela
revolu\cao da inform\'atica. Isso continuar\'a ocorrendo sempre que a
capacidade de tratar informa\cao aumente. No entanto, o crecimento 
da rapidez das computadoras est\'a associada a uma maior capacidade de
miniaturiza\c c\~ao. Aqui vale lembrar que o ponto de partida de tudo foi 
a descoberta do efeito transistor~\cite{adventures}. Mais ainda, o primeiro
transistor tinha dimens\~oes macrosc\'opicas e seu pre\c co era
da ordem de USA\$ 1. Dai para c\'a por esse pre\c co podem-se comprar
milh\~oes deles! Foi isso que permitiu a revolu\c c\~ao da inform\'atica
n\ao prevista mesmo por von Neumann (veja discus\~ao na Sec.~\ref{sec:x}).  

De fato a densidade de transistores em cada chip aumentou 
exponencialmente nos \'ultimos 24 anos. 
Manter esse ritmo nos pr\'oximos anos implicar\'a em confrontar, em algun 
momento, as barreiras da mec\^anica qu\^antica. 
Toda a ci\^encia e a tecnologia nanom\'etrica \'e dominada pelos efeitos 
qu\^anticos. A escrita de dimens\~oes nanom\'etricas est\'a cada vez mais
desenvolvida~\cite{hong}; as suas aplica\coes v\ao desde a qu\ii mica (onde 
as t\'ecnicas poderiam ser usadas para controlar a dist\^ancia entre
os reagentes numa rea\cao qu\ii mica) at\'e dispositivos eletr\^onicos
com dimens\~oes moleculares. A ci\^encia aplicada chega cada vez mais perto
da ci\^encia b\'asica. Neste dom\ii nios  os fen\^omenos qu\^anticos ser\ao 
cada vez mais importantes. 

Assim entender melhor essa estranha e bela teoria ser\'a um dos mais 
importantes temas que a f\ii sica vai brindar ao resto das cie\^encias e,
em geral, a todas as outras formas das atividades humanas. 
Por outro lado e n\ao menos espetacular ser\'a o 
controle da computa\cao qu\^antica~\cite{jp}. 

A maioria das \'areas do conhecimento puderam ter 
grande desenvolvimento nas ultimos anos
apenas pelos avan\c cos na inform\'atica e esta contou e continuar\'a a
contar, direta ou indiretamente, com a participa\cao dos f\ii sicos. 
\'E pos isso que muitos f\ii sicos continuar\ao a fazer f\ii sica e muitos 
biol\'ogos passar\ao a  pensar cada vez mais ... como f\ii sicos! 
A  f\ii sica est\'a longe de estar esgotada~\cite{ld}.

\section{S\ii ntese versus diversidade}
\label{sec:sinte}

Algumas vezes a f\ii sica se encontra em situa\coes de s\ii ntese,
enquanto na maior parte das vezes \'e a diversidade a que prevalece. De
fato, a diversidade \'e uma carater\ii stica das ci\^encias
desenvolvidas. 

Segundo Dyson~\cite{dyson} por per\ii odos longos as diversas ci\^encias
permanecem dominadas pela {\em concretitude}. Por exemplo, na maior parte 
do s\'eculo XIX e nas d\'ecadas posteriores aos anos 30 deste s\'eculo. 
Em outras ocasi\~oes, \'e a {\em abstra\cao} que domina. Os pesquisadores 
de uma \'epoca determinada \nao podem escolher entre qual a tend\^encia que
domine. Isso est\'a definido por fatores externos e, muitas vezes
pelo acaso. Depois das revolu\coes da mec\^anica qu\^antica e
relatividade restrita e geral, como poderiamos esperar o desenvolvimento de
esquemas te\'oricos mais gerais ainda num breve per\ii odo de tempo? 

No entanto, progressos 
importantes foram conseguidos no per\ii odo de 1960-1980. O chamado
{\em modelo padr\ao} das intera\coes eletrofracas e fortes
foi uma conquista do ponto de vista da teoria qu\^antica de
campos, acrescentadas de descobertas te\'oricas como a {\em liberdade
assint\^otica} e o {\em mecanismo de Higgs} mencionados antes. A
partir da\'\i, uma s\'erie de extrapola\coes dessas 
estruturas levaram \`a maioria dos f\ii sicos a pensar que a s\ii ntese
final estaria chegando ao fim. Isso fica evidente nas palavras de Hawking 
citadas an\-te\-ri\-or\-men\-te.

A vi\sao de Dirac dominou a f\ii sica nas d\'ecadas passadas.
Em 1970 L\'eon van Hove dizia~\cite{sss} 
\begin{quotation}
{\em ... physics now look more like chemistry in the sense that...a much
larger fraction of the total research deals with complex systems,
structure and processes, as against a smaller fraction concerned with
the fundamental laws of motion and interactions...we all believe that
the fundamentals of classical mechanics, of the electromagnetic
interaction, and of statistical mechanics dominate the multifarious
transitions and phenomena you discuss this week;
and I assume
that none of you expects his work on such problem to lead to
modifications of these laws. You known the equations more than the
phenomena... }
\end{quotation}

Agora sabemos que \nao foi bem assim. Novas leis fundamentais estavam
sendo descobertas pelos te\'oricos e em pouco tempo testadas pelos
f\ii sicos experimentais. (Lembremos do caos na mec\^anica cl\'assica.) 
No fundo, temos a esperanza que isso
aconte\c ca de novo. \'E, no entanto, pouco prov\'avel a curto prazo. 
No paradigma das teorias de grande unifica\c c\~ao, supersimetria e
supercordas o problema \'e que as predi\coes \nao ambiguas destas teorias 
s\'o ocorrem a escalas que dificilmente ser\ao atingidas pela f\ii sica
experimental em curto prazo. 
A possibilidade seria uma mudan\c ca de paradigma. Por\'em, \nao h\'a 
nenhuma proposta te\'orica que traga uma luz nesse sentido. Mas, como 
dissemos antes, a natureza \'e mais esperta do que n\'os. 

A ci\^encia progride lentamente sem se importar com nossas pressas e 
angustias. Em 1896, ou seja antes da descoberta do n\'ucleo at\^omico e
da mec\^anica qu\^antica, Emil Wiechert disse~\cite{dyson}
\begin{quotation}
{\em A mat\'eria que supomos ser o principal componente do universo \'e
formada por tijolos independentes, os \'atomos qu\ii micos. Nunca
ser\'a demais repetir que a palavra ``\'atomo'' est\'a hoje em dia
separada de qualquer especula\cao filos\'ofica antiga: sabemos
precisamente que os \'atomos com os quais estamos lidando n\ao s\ao
em nenhum sentido os mais simples componentes conceb\ii veis do
universo. Ao contr\'ario, diversos fen\^omenos, especialmente na
\'area da espectroscopia, levam \`a conclus\ao de que os \'atomos
\sao estruturas bastante complexas. At\'e onde vai a ci\^encia
moderna, devemos abandonar por completo a id\'eia de que 
penetrando no limiar do pequeno conseguiremos alcan\c car as
funda\coes finais do universo. Acredito que podemos abandonar essa
id\'eia sem nenhum remorso. O universo \'e infinito em todas as
dire\c c\~oes, \nao apenas acima de n\'os, na grandeza, mas tamb\'em
abaixo de n\'os, na pequenhez. Se partirmos da nossa escala humana de
exist\^encia e explorarmos o conte\'udo do universo al\'em e al\'em,
chegaremos finalmente, tanto no reino do pequeno quanto no reino do
grande, a dist\^ancias obscuras onde primeiro nos nossos sentidos e
depois nossos conceitos nos falhar\~ao.}
\end{quotation}
Depois disso fica dif\ii cil entender as palavras do Mach acima!
Definitivamente os f\ii sicos hoje  em dia n\ao pensam mais como Mach.
Por exemplo, \'e interessante a posi\cao de Dyson~\cite{dyson}:
\begin{quotation}
{\em ...a Natureza \'e complexa. J\'a \nao \'e mais nossa a vis\ao que
Einstein conservaria at\'e sua morte, de um mundo objetivo de espa\c
co, tempo e mat\'eria, independente do pensamento e da observa\cao
humanos. Einstein esperava encontrar um universo dotado do que
chamava `realidade objetiva', de um universo de picos montanhosos
que ele poderia compreender por meio de um conjunto finito de equa\c
c\~oes. A natureza, como em fim se descobriu, vive \nao nos cumes
elevados, mas nos vales l\'a embaixo.}
\end{quotation}
Este tipo de posicionamento ainda que mais frequentes na atualidade, 
\nao \sao majorit\'arias.

Sabe-se agora que h\'a milhares de teorias de supercordas que \sao 
ma\-te\-ma\-ti\-ca\-men\-te consistentes da mesma maneira que as duas 
teorias de Green e Schwarz. Esta consist\^encia matem\'atica \'e 
garantida pela invari\^ancia conforme.   
A menos que seja mostrado que essa diversidade de teorias \sao 
e\-qui\-va\-len\-tes, 
a \'unica e importante consequ\^encia das teorias de supercordas \'e que 
as simetrias do espa\c co-tempo e internas \nao \sao colocadas a m\~ao. 
De fato, em 1985 j\'a se tinham reduzido a 5 as teorias de supercordas 
diferentes. Logo depois, a introdu\c c\~ao de um novo tipo de simetria 
chamada de {\it dualidade-S} (o exemplo cl\'assico \'e a dualidade dos 
campos el\'etrico e magn\'etico) reduz a apenas o n\'umero a 3. Mais 
recentemente, com a descoberta de novas dualidades as 5 de supercordas 
em 10 dmens\~oes e 
uma teoria de campos em 11  dimens\~oes s\ao consideradas apenas
a manifesta\cao  de apenas uma teoria--$M$, ainda que n\ao exista uma
formula\cao completa desse tipo de teoria~\cite{nathan,duff,hilg}.
O problema \'e que `` no one knows how to write down the equation of this 
theory''~\cite{sw3}.
De qualquer forma, o limite de baixas energias tem de ser escolhido antes. 
Por exemplo, se for confirmado um modelo que inclua o modelo padr\ao 
(de maneira unificada ou n\~ao) ent\ao deve haver uma teoria de 
supercordas cujo limite se baixas energias seja esse modelo e \nao outro. 
(Um das teorias de Green e Schwarz tem um limite de baixas energias perto 
do modelo padr\~ao. Mas at\'e o momento \nao foi encontrada uma teoria que 
reprodu\c ca a baixas energias os quarks e leptons conhecidos.)  Mesmo que 
algu\'em descobrisse qual \'e essa teoria de supercordas, \nao saberiamos 
explicar porque esta teoria \'e a que descreve o mundo real. 
O objetivo da f\ii sica \nao \'e apenas descrever o mundo mas explicar 
porque ele \'e como \'e~\cite{sw2}.

Devemos por tanto dar maior import\^ancia aos detalhes. 
Por exemplo, \nao \'e qualquer conjunto de equa\coes diferenciais parciais 
que descreve o campo eletromagn\'etico. S\ao apenas as equa\coes de Maxwell 
que o fazem. Da mesma maneira \nao \'e qualquer teoria n\~ao-Abeliana que 
descreve as intera\coes de quarks e leptons. 
Em ambos casos, as equa\coes de Maxwell e o modelo padr\~ao, sempre 
podemos estudar maneiras de generaliz\'a-los. Mas, \nao ser\'a qualquer 
generaliza\cao que ser\'a seguida pela natureza.

Mesmo que dispuss\'essemos de uma teoria de supercordas real\ii stica, 
que explicasse tudo o que modelo padr\ao deixa em aberto (as massas dos 
fermions por exemplo) ainda teriamos que explicar porque essa teoria e 
suas asun\coes importantes \sao escolhidas pela natureza. Em outras 
palavras, \'e mais 
prov\'avel que essa teoria de supercordas precisse de princ\ii pios ainda 
mais profundos para ser explicada.

Usualmente Einstein \'e considerado como um dos defensores da procura de 
leis unificadas da natureza. No entanto, ele via isso, pelo menos num 
per\ii odo da sua vida, como um processo sem fim. Em 1917 ele escreveu 
para Felix Klein~\cite{pais1}
\begin{quotation}
{\em However we select from nature a complex [of phenomena] using the
criterion of simplicity, in no case will its theoretical treatment
turn out to be forever appropriate (sufficient). Newtons's theory,
for example, represent the gravitational field in a seemingly
complete way by means of the potential $\phi$. This description
proves to be wanting; the functions $g_{\mu\nu}$ take its place. But
I do not doubt that the day will come when that description, too,
will have to yield to another one, for reason which at present we do
not yet surmise. I belive that this process of deeping the theory has
no limits.}
\end{quotation}

Klein escreveu para Einstein nesse mesmo ano falando sobre a invari\^ancia 
conforme na eletrodin\^amica. Einstein respondeu~\cite{pais1}
\begin{quotation}
{\em It does seem to me that you highly overrate the value of formal
point of view. These may be valuable when an \verb+ already found+
truth needs to be formulated in a final form, but fail
almost always as heuristics aids.}
\end{quotation}

Segundo Pais~\cite{pais1}
\begin{quotation}
{\em Nothing is more striking about the later Einstein than his change
of position in regard to this advice, give when he was in his late
thirties.}
\end{quotation}

Quer dizer que segundo esta vi\sao inicial de Einstein, a cadeia de 
explica\coes em termos de princ\ii pios cada vez mais gerais e profundos, 
\nao teria fim. Assim, uma teoria de tudo \nao seria poss\ii vel. Por 
outro lado, 
segundo Weinberg~\cite{sw2}, o fato que nossos princ\ii pios t\^em-se 
tornado mais simples e econ\^omicos, poderia indicar que {\sl deva} 
haber uma tal teoria. No entanto, o que queremos enfatizar aqui \nao 
\'e se concordamos ou 
\nao com as posi\coes do tipo das de Weinberg. 
Queremos \'e colocar se a nossa compreen\sao da 
estrutura \ii ntima da mat\'eria ser\'a, em \'ultima inst\^ancia, 
melhorada se dedicarmos mais esfor\c cos \`a f\ii sica de 1 TeV ou \`a 
da escala de 
Planck ($10^{19}$ GeV) ou, mesmo se novas leis ``macrosc\'opicas'' 
poder\~ao ser de utilidade na compreens\~ao \'ultima do universo. 

Uma ``teoria de tudo'' \nao seria eficaz com rela\cao ao problema da 
complexidade organizada existente na natureza. De fato, \'e poss\ii vel 
que as leis que regem a complexidade e que seriam v\'alidas para qualquer 
sistema complexo, incluindo o universo, \nao sejam do tipo das leis da 
natureza conhecidas at\'e o momento~\cite{jb}. 

Pesquisar os detalhes de ``vales e montanhas'', para usar a analogia de
Deyson,  pode ser mais 
interessante que pesquisar os  picos. No m\ii nimo vai se encontrar coisas
diferentes e n\~ao menos importantes. Pior, \nao temos escolha. Se 
precissamos pesquisar detalhes ou n\~ao, depende do desenvolvimento de uma 
determinada ci\^encia. Os vales \nao \sao apenas um ponto de refer\^encia 
para medir a altitude dos picos. Eles t\^em a sua pr\'opria diversidade,
suas pr\'oprias leis e metodologia. Suas pr\'oprias surpressas.
N\ao devemos ter medo de que a f\ii sica, pelo momento, se torne uma 
bot\^anica ou uma qu\ii mica. 
Depois, uma nova ordem geral vir\'a. De novo podemos chamar a 
aten\cao aqui para o tect\^onica de placas. Por muitos s\'eculos os estudos
da Terra eram ``chatos''. Procurava-se classificar as rochas! mas sem essa
fase aborrecida n\ao teriamos a s\ii ntese atual.

Apenas agindo podemos ver o que realmente acontecer\'a.
Devemos {\em participar} do processo cient\ii fico. \'E por
isso que a decis\ao pessoal mencionada acima \'e importante. N\ao
\'e apenas uma quest\ao de opini\~ao. Segundo a decis\ao tomada seguiremos
um ou outro caminho na nossa pesquisa e segundo esse caminho poderemos
ser melhor ou pior sucedidos. Melhor, se a\-cre\-di\-ta\-mos que o processo 
cient\ii fico, longe de acabar, est\'a apenas come\c cando devemos 
tornar-nos participes dele o quanto antes.

Se ainda \'e o sonho de alguns f\ii sicos a formula\cao de uma teoria que
unifique todas as intera\coes conhecidas~\cite{hgeorgi2}, 
devemos ter sempre presente que uma teoria de tudo \'e em princ\ii pio
imposs\ii vel. Toda teoria tem sua componente fenomenologico,
aspectos que \nao podem ser calcul\'aveis usando os conceitos da
mesma teoria. Todas as teorias \sao e ser\ao aproximadas, estaremos
sempre numa ``unended quest''. E isso \'e empolgante.

Nos \'ultimos anos a nossa compreens\ao da teoria qu\^antica de campos
mudou consideravelmente. A  descri\cao das part\ii culas em teoria 
qu\^antica
de campos depende da energia na qual estudamos essas intera\c c\~oes. 
Assim todas as teorias podem ser consideradas como {\em teorias efetivas}, 
levando em conta apenas as part\ii culas relevantes na escala de energia 
considerada.
 
Este \'e uma realiza\cao do fato que podemos estudar fen\^omenos ou
processos f\ii sicos apenas num intervalo limitado de energia.
Aparecem infinitos pela exig\^encia de {\em localidade} que significa
que a cria\cao e/ou aniquila\cao de part\ii culas ocorre num ponto do 
espa\c co-tempo. O processo de {\em renormaliza\c c\~ao} foi interpretado
at\'e recentemente como uma maneira de absorver os infinitos nos 
par\^ametros f\ii sicos. Isto \'e, introduzindo um {\sl cut-off} $\Lambda$ 
e modificando a teoria para dist\^ancias menores que 
$\Lambda^{-1}$, aparecem apenas quantidades pass\ii veis de serem 
medidas experimentalmente como a massa e a carga el\'etrica. 
Finalmente faz-se $\Lambda\to\infty$.
A teoria fica independente do {\sl cut-off}. Segundo Dirac, quem nunca 
aceitou o processo de renormaliza\c c\~ao, a f\ii sica te\'orica tomou 
um pista errada com esse desenvolvimento. Hoje em dia essa vis\ao 
(de Dirac) \'e considerada muito restrita. Afinal a eletrodin\^amica 
qu\^antica \'e apenas uma parte do modelo eletrofraco que certamente 
\'e  parte de uma
teoria mais abrangente. O problemas dos infinitos ser\'a resolvido 
quando tivermos uma teoria final (se \'e que ela existe). Este caso 
\'e interessante para refletir. As vezes os problemas apenas existem 
porque estamos supervalorizando nossos objetos de estudo. Por exemplo, 
Kepler prop\^os 
um modelo das \'orbitas dos planetas baseado em simetrias. Agora sabemos
que as simetrias fundamentais \nao aparecem nesse tipo de sistemas. Assim
a proposta de Kepler era interessante mais aplicada no problema errado.
Os planetas \nao \sao mais objetos fundamentais para a formula\cao 
de teorias f\ii sicas b\'asicas. Esta discu\sao parece vanal mas n\ao \'e.
Muitas escolhas de estudantes ou mesmo de pesquisadores ser\ao feitas
segundo o aspecto que valorizem na pesquisa. Isto \'e, o que seja 
considerado um problema importante.  Se algu\'em concorda com Dirac que 
``a eletrodin\^amica qu\^antica atual \nao corresponde ao elevado padr\ao
de beleza matem\'atica que seria de esperar de uma teoria f\ii sica 
fundamental'' e tentar modificar {\em apenas } a QED poder\'a  encontrar
problemas insoluv\'eis mesmo para mentes bem preparadas e privilegiadas.

As teorias de grande unifica\cao tentavam esclarecer melhor o modelo
padr\~ao. Por exemplo dar resposta ao problema da quantiza\cao da carga
e tentar calcular o \^angulo de mistura eletrofraco ($\sin\theta_W$).
Certamente n\ao foram propostas como continua\cao das ideias de 
Einstein~\cite{hgeorgi2}. Segundo Georgi
\begin{quotation}
{\em Einstein's attempts at unification were rearguard action which 
ignored the
real physics of quantum mechanical interactions between particles in the 
name of philosophical and mathematical e\-le\-gance. Unfortunately, it 
seems to me that many of my colleages are repeating the Eintein's mistake.

The progress of the fields is determined, in the long run, by the
progress of experimental physics. Theorists are, after all, 
pa\-ra\-si\-tes.
Without our experimental friends to do the real work, we might as
 well be mathematicians or philosophers. When the s\-cien\-ce is healthy, 
theoretical
and experimental particle physics track along together, each reforcing
the other. But there are often short period during which one or other 
aspect of the field gets away ahead. Then theorists tend to lose contact 
with reality
...During such periods without experiments to exited them, theorists tend 
to relax back into their grounds states, each doing whatever come most 
naturally. As a result, since different theorists have different skills,
the field tends to fragment into little subfields. Finally, when the
crucial ideas or the crucial experiments come along and the field regains
its vitality, most theorists find that they have been doing irrelevants
things...But the wonderful thing about physics is that good theorists
don't keep doing irrelevant things after experiment has spoken. The useless
subfields are pruned away and everyone does more or less the same thing
for a  while, until the next boring period.}
\end{quotation}

Segundo Heisenberg~\cite{wh}, em f\ii sica te\'orica podem-se, i) formular
teorias fenomenol\'ogicas, ii) esquemas matem\'aticos rigorosos, ou iii)
usar a fi\-lo\-so\-fia como guia. 
No primeiro tipo ficam as pesquisas de Heisenberg
e o grupo de Sommerfeld (Pauli, Land\'e, H\"oln e outros). Eles inventavam
f\'ormulas que reproduzissem os experimentos. No entanto, mesmo quando
bem sucedidas essa teorias fenomenol\'ogicas \nao fornecem nenhuma 
informa\cao real sobre o conte\'udo f\ii sico do fen\^omeno~\cite{wh}. As
vezes os c\'alculos feitos com ambos esquemas coincidem. Isto \nao \'e
de se estranhar. A equival\^encia pode ser matem\'atica mas \nao f\ii sica.

No entanto, as vezes resultados rigorosos tamb\'em levam a resultados
em discord\^ancia do observado. Isto \'e, eles t\^em tamb\'em 
limita\c c\~oes. A tese de doutorado de Heisenberg versou sobre o c\'alculo
da estabilidade de um fluxo entre duas paredes fixas. O resultado foi que 
para um certo n\'umero de Reynolds o fluxo torna-se inst\'avel e turbulento.
Um ano depois E. Noether mostrou
rigorosamente que o problema de Heisenberg \nao tinha solu\c c\~ao: o 
fluxo devia ser est\'avel em toda parte. Outro exemplo mais recente sobre
as limita\coes das demostra\coes gerais \'e o do mecanismo de Higgs, que
nada mais \'e do que uma evas\~ao do teorema de Goldstone.
Este fora recebido com descren\c ca por muitos te\'oricos pois o teorema 
de Goldstone tinha sido mostrado rigorosamente pelos axiom\'aticos.  Antes 
do seminario que daria em Princeton, Higgs conta que segundo Klaus Hepp
\begin{quotation}
{\em ... what I going to
say must be nonsense because axiomatic field theorist had proved
the Goldstone theorem rigorously using the methos of $C^*$-algebras.
However, I survive questions from Arthur Wightman and others, so I
conclude that perhaps the $C^*$ algebraists should look again}~\cite{ph}.
\end{quotation}
Sem coment\'arios. 

Heisenberg disse que nunca se soube onde estava o erro na demostra\cao de 
Noether mas em 1944 (vinte anos depois) Dryden e colaboradores fizeram 
experi\^encias precissas do fluxo laminar entre duas paredes e da 
transi\cao para a turbul\^encia e descobriram que os c\'alculos de 
Heisenbeg estavam
em concord\^ancia com a experi\^encia. Lin do MIT simulou a experi\^encia 
(von Neumann sugeriu usar computadores) e confirmou de novo os resultados 
de Heisenberg. Poderiamos
colocar outros exemplos, mas esses \sao suficientes para mostrar as 
limita\coes dos m\'etodos matem\'aticos rigurosos. Devemos lembrar,
para fazer justi\c ca que com os m\'etodos fenomenol\'ogicos somos
obrigados a usar sempre os velhos conceitos mesmo para uma situa\cao nova.
Bom, isto quer dizer que qualquer que seja o m\'etodo usado na pesquisa
te\'orica apenas quando verificado experimentalmente~\footnote{Aqui essa
``verifica\cao~'' \'e entendida de maneira ampla. Pode ser apenas
indireta, por exemplo, uma consist\^encia global da teoria com os dados
experimentais. Isto \'e o que ocorre com o modelo padr\ao da f\ii sica
das part\ii culas elementares.} podemos dizer que a teoria funcionou.
Tamb\'em devemos enfatizar que o passo decisivo \'e sempre discont\ii nuo. 
Isso aconteceu, por exemplo, com a mec\^anica qu\^antica~\cite{wh}.  

A filosofia como guia da pesquisa te\'orica teve seu apogeu com o 
po\-si\-ti\-vis\-mo. Mach, e depois o circulo de Viena (recentemente 
te\'oricos como Chew), insistiram que uma teoria devia ser formulada em 
termos de quantidades observ\'aveis. Mach, e muitos dos seus 
contempor\'aneos, acreditava que os \'atomos eram apenas uma quest\ao de 
conveni\^encia, \nao creditavam na exist\^encia real deles.  
Einstein fora influenciado por esta vi\sao no 
come\c co da sua carreira mas logo mudou de id\'eia. Ele diria que~\cite{wh} 
\begin{quotation}
{\em ... a posibilidade que se tem de observar ou \nao uma coisa depende da 
teoria que se usa. \'E a teoria que decide o que pode ou \nao ser 
observado}.
\end{quotation}
 
\section{Conclus\~oes}
\label{con}                                               

Pode a comunidade cient\ii fica errar o rumo? \'E uma quest\ao delicada.
Uma teoria \'e aceita pela comunidade dependendo de v\'arios fatores como:
sua e\-xa\-ti\-d\ao nas predi\c c\~oes, seu contexto e o grau em que est\'a
determinada pela experi\^encia. Por\'em, a curto prazo, existem outros 
fatores que \nao \'e f\'acil de reconhecer como sendo esp\'urios: modismo,
ideologia e estupidez ge\-ne\-ra\-li\-za\-da. 
As vezes o prazo \nao \'e t\ao curto
assim: o modelo solar de Aristarco passou despercevido pelos astron\^omos
ao longo de 17 s\'eculos e por quase 200 anos a teoria da via L\'actea
de Kant tampouco foi popular nos meios acad\^emicos. 
Recentemente alguns aspecto da mec\^anica qu\^antica est\~ao sendo 
esclarecidos, as conclus\~oes de N. Bohr seriam corretas mas pelos 
argumentos 
errados?~\cite{durr}.

Deve-se insistir com os estudantes que um aspecto que importa (certamente 
\nao o \'unico) no que-fazer cient\ii fico \'e a ``emo\c c\~ao'', qualquer 
coisa que isso signifique. Segundo  Kadanoff~\cite{ruelle}
\begin{quotation}
{\em \'E uma experi\^encia como nenhuma outra que eu possa des\-cre\-ver; a
melhor coisa que pode acontecer a um cientista, compreender que
alguma coisa que ocorreu em sua mente corresponde e\-xa\-ta\-men\-te a 
alguma coisa que acontece na natureza. \'E surpreendente, todas as vezes que
ocorre. Ficamos espantados com o fato de que um construto de nossa
pr\'opria mente possa realmente materializar-se  no mundo real que
existe l\'a fora. Um grande choque e uma alegria muito grande.}
\end{quotation}
ou, de maneira mais dramatica nas palavras de Einstein
\begin{quotation}
{\em The years of anxious searching in the dark, with their intense longing, 
their alternations of confidence and exhaustion and the final emegence into 
the light---only those who have experienced it can understand it.}
\end{quotation}
Para sentir essa emo\coes \nao precissamos obter resultados t\ao importantes 
quanto os de Kadanoff e Einstein! Apenas devem ser resultados
{\it nossos}.

Nos Estados Unidos os estudantes est\ao deixando a academia para trabalhar
na empressa privada. Isso n\ao seria problema se entre eles, segundo 
Anderson~\cite{pwa2}, n\ao estivessem os melhores. Os menos criativos
ficam nas posi\coes permanentes em f\ii sica. Segundo Anderson a National 
Science Foundation (NSF) e outras ag\^encias de fomento est\ao incentivando
a falta de criatividade, talvez, influenciados pelo 
``Horganism''.~\footnote{A cren\c ca que o fim da ci\^encia est\'a 
pr\'oximo e, o que fica \'e
apenas per\'\i odos de ``ci\^encia normal'' segundo a vis\ao de 
Kunh~\cite{tk}.} As causas disso reside pelo menos em parte tamb\'em no 
sistema de ``peer-review'' mas este \'e um aspecto que n\~ao vai ser 
discutido neste artigo, serve apenas como uma confirma\cao de que 
a vis\ao pessoal que temos sobre
o que-fazer cient\ii fico e o futuro de ci\^encia t\^em implica\coes no 
desenvolvimento da pr\'opria ci\^encia.
  
Uma carater\ii stica do nosso tempo \'e a pressa. N\ao apenas na ci\^encia.
Umberto Eco trata do problema da rapidez~\cite{eco}
\begin{quotation}
{\it Quando enalteceu a rapidez, Calvino preveniu: `N\ao quero
dizer que a rapidez \'e um valor em si. O tempo narrativo pode
ser lento, c\'\i clico ou im\'ovel...Esta apologia da rapidez \nao 
pretende negar os prazeres da demora. Se algo importante ou
absorvente est\'a ocorrendo, temos de cultivar a arte da demora.}
\end{quotation}
Isto \'e v\'alido \nao s\'o na fic\cao mas tamb\'em na ci\^encia.

Como podemos ser pesimistas se nos \'ultimos anos a f\ii sica foi capaz
de i) encontrar algumas leis da natureza novas, ii) obter novos
comportamentos da natureza, iii) desenvolver instrumentos que permitiram
observar fen\^omenos em condi\coes completamente diferentes das estudadas
no pasado?~\cite{vw1}. As experi\^encias em Stanford no fim dos anos
60, e que desvendaram a estrutura do n\'ucleob, 
\nao foram meras repeti\coes da experi\^encia de Geiger-Marsden.
O conceito de visualiza\cao dos fen\^omenos tinha mudado. De fato
novas maneiras de estudar (``ver'') a natureza \sao t\ao importantes quanto
as leis fundamentais que surgir\ao desses estudos.

\begin{quotation}
{\em  Chegamos ao fim?...que direito temos de supor que os nucleons,
el\'etrons e neutrinos \sao realmente elementares e \nao podem 
ser subdivididos em pares constituintes ainda menores? H\'a apenas meio 
s\'eculo, \nao se supunha que os \'atomos eram indivis\ii veis?...
embora seja imposs\ii vel prever o desenvolvimento futuro da ci\^encia da
mat\'eria, temos atualmente raz\~oes para acreditar que nossas
part\ii culas elementares \sao na verdade as unidades b\'asicas e \nao
podem ser novamente subdivididas...parece, assim, que chegamos ao fim de
nossa pergunta dos elementos b\'asicos que formam a mat\'eria.}
\end{quotation}
Estas palavras foram escritas por George Gamow em 1960~\cite{gamow}.
\'E interessante que elas continuem sendo, em parte verdadeiras.
Os nucleons, el\'etrons e neutrinos continuam a ser indivis\ii veis
no sentido direto. Como podia imaginar Gamow que os nucleons seriam
divis\ii veis ``em certo sentido''? Somos capazes de estudar a estrutura
dos nucleons mas seus constituentes est\~ao, aparentemente, confinados!

Segundo Weinberg, 
\begin{quotation}
{\em ... it is foolhardy to assume that one knows even the terms in which a 
future final theory will be formulated.}
\end{quotation}

\'E dif\ii cil usar argumentos gerais sobre a utopia que serviria
de guia para o avan\c co da f\ii sica de altas energias. Segundo 
Bohr~\cite{wh}
\begin{quotation}
{\em Quando se tem uma formula\cao correta, o  oposto dela \'e, 
e\-vi\-den\-te\-men\-te
uma formula\cao errada. Mas quando se tem uma verdade profunda, ent\ao seu
oposto pode ser igualmente uma verdade profunda}.
\end{quotation}
Assim, assumir que devem existir leis gerais com as
quais possam ser des\-cri\-tas, pelo menos em princ\ii pio, todas as coisas
parece uma verdade profunda. O seu oposto, que no fundo \nao existem leis 
fundamentais tamb\'em o \'e (como arg\"uem Wheeler e Nielsen~\cite{sw2}).

Desde 1859 sabia-se que havia um problema com a \'orbita de mercurio se 
interpretada dentro da teoria da gravita\cao de Newton. Este problema, 
como \'e bem conhecido, foi resolvido pela teoria da relatividade geral de 
Einstein. Mas, em 1916 al\'em dessa discrep\^ancia haviam tamb\'em mais duas.
Uma referia-se a anomalias relativas aos movimentos dos cometas Halley e 
Encke. A outra era a respeito do movimento da lua~\cite{sw2}. Em todos estes 
casos, como no da \'orbita de mercurio, os movimentos \nao concordavam com 
as previs\~oes da teoria de Newton. 
Agora, no entanto, sabe-se que as anomalias nos 
movimentos dos cometas \sao devidas \`a press\ao de escape dos gases ja que 
o cometa \'e esquentado quando passa perto do sol. O movimento da lua foi 
melhor comprendido quando se levou em conta o seu tamanho que implica em 
complicadas for\c cas tidais. Assim, segundo Weinberg~\cite{sw2} 
\begin{quotation}
{\em ...there is nothing in any single 
disagreement
 between theory and experiment that stands up and waves a flag and says 
`` I am an important anomaly''}.
\end{quotation} 
Assim, \nao sabemos em geral quando estamos lidando com um verdadeiro sinal 
de f\ii sica nova. 

Usualmente a maneira de fazer f\ii sica de Dirac \'e considerada com
a maneira matem\'atica. Mas ele tinha uma posi\cao mais ampla e um
conhecimento das limita\coes dessa maneira de trabalhar~\cite{wh}
\begin{quotation}
{\em Em qualquer parte da f\ii sica em que se saiba muito pouco, somos 
obrigados a nos prender \`a base experimental, sob pena de mergulharmos em 
especula\coes estravagantes, que quase certamente estar\ao erradas. N\ao
desejo condenar completamente a es\-pe\-cu\-la\-\c c\~ao. 
Ela pode ser divertida e indiretamente \'util, mesmo que acabe por se 
mostrar errada...mas \'e precisso tomar cuidado  para \nao se deixar 
envolver demais por ela.}
\end{quotation}

Onde fica a intui\c c\~ao em tudo isto? quais os limites do m\'etodo 
cient\ii fico? Vale a pena se preocupar com isto? as respostas s\~ao 
pessoais. 
Um exemplo da import\^ancia da emo\c c\~ao \'e expressa por Thomas Mann 
quando escreveu 
\begin{quotation}
{\em  Astronomy---a great science---teaches us to consider the earth as a 
comparison of an insignificant star in the giant cosmic turnoil, roving 
about at the the periphery of our galaxy.
This is, no doubt, correct. But I doubt that such correctness reveals
the whole truth. In the depth of my soul I belive---that this earth has a 
central significance in the universe. In the depth of my soul I entertain
the presumption that the act of creation which called forth the inorganic 
world, from nothingness, and the procreation of life from the inorganic
world, was aimed at humanity. A greart experiment was initiated, whose
failure by human irresponsability would mean the failure of the act of
creation itself, its very refutation. May be it is so, mat be it is not.
It would be good if humanity behaved as if it were.}
\end{quotation}

As vezes os artistas enxergam mais longe que os
cientistas. Um deles ja disse, s\'eculos atr\'as
\begin{quotation}
{\em There are more things in Heaven and Earth. Horatio.\par
Than are dreamt of in your philosophy. }
\end{quotation}

Precissamos convencer os estudantes que existem (e que sempre existir\~ao) 
muitas coisas a serem des\-co\-ber\-tas, talvez virando \`a esquina. 
Convenc\^e-los que o progresso cient\'\i fico e tecnol\'ogico
foi obtido lentamente, e por vezes de maneira ca\'otica, e que \nao
existe uma raz\~ao para que surpressas n\ao ocorram de novo. Que a 
pressa n\~ao serve para queimar etapas. 
Que previs\~oes s\~ao dif\'\i ceis de se fazer. N\ao apenas para n\'os mas 
que tamb\'em era dif\'\i cil para von Neumann. 
Precisamos colocar em discus\~ao a maneira como se processa
o desenvolvimento das ideias cient\ii fica, seus acasos, atrasos e 
acelera\coes devido a preconceitos que n\ao fazem parte do m\'etodo 
cient\'\i fico, mas est\ao sempre presente para bem ou para mal. 
Isso implica a valoriza\cao da perpectiva hist\'orica no ensino de 
ci\^encias.

\vskip 2cm
\begin{center}
\vskip 2cm
{\bf Agradecimentos}
\end{center}

Agrade\c co ao CNPq pelo auxilio financeiro parcial.

\end{document}